\documentclass[onecolumn,showpacs,preprintnumbers, 
               superscriptaddress,eqsecnum,nofootinbib]{revtex4}
\usepackage{graphicx, fancybox}
\usepackage{amsmath,amssymb,color}
\usepackage[dvipdfm, colorlinks=true, pdfstartview=FitV, linkcolor=red,
 citecolor=blue, urlcolor=blue]{hyperref}

\newcommand{\vect}[1]{\mathbf{#1}}

\newcommand{\im}{\mathrm{Im}}
\newcommand{\re}{\mathrm{Re}}
\newcommand{\sgn}{\mathrm{sgn}}

\newcommand{\Slash}[1]{\ooalign{\hfil/\hfil\crcr$#1$}}

\newcommand{\vz}{\vect{0}}
\newcommand{\vp}{\vect{p}}
\newcommand{\vk}{\vect{k}}
\newcommand{\vgamma}{{\boldsymbol \gamma}}
\newcommand{\Tr}{\mathrm{Tr}}
\newcommand{\pw}{(\vp,\omega)}

\newcommand{\Tnonzero}{_{T\neq 0}}

\newcommand{\imagm}{(\vp,i\omega_m;m)}
\newcommand{\imagalpham}{(\vp,i\omega_m;\sqrt{\alpha}m)}
\newcommand{\Bhard}{{\text{Re}}\tilde{B}(\vz,\omega;m)^{E_f>T^*}\Tnonzero}
\newcommand{\Bzerohard}{{\text{Re}}\tilde{B}^0(\vz,\omega;m)^{E_f>T^*}\Tnonzero}
\newcommand{\Bsoft}{{\text{Re}}\tilde{B}(\vz,\omega;m)^{E_f<T^*}\Tnonzero}
\newcommand{\Bzerosoft}{{\text{Re}}\tilde{B}^0(\vz,\omega;m)^{E_f<T^*}\Tnonzero}
\newcommand{\sig}[1]{C_{#1}(\omega)}
\newcommand{\one}{\sig{1}}
\newcommand{\two}{\sig{2}}
\newcommand{\three}{\sig{3}}
\newcommand{\four}{\sig{4}}
\newcommand{\five}{\sig{5}}
\newcommand{\six}{\sig{6}}
\newcommand{\seven}{\sig{7}}
\newcommand{\mOvergT}{\lambda}

\preprint{KUNS-2315}

\begin{document}

\title{Spectral Function of Fermion Coupled with Massive Vector Boson at Finite Temperature in Gauge Invariant Formalism}

\author{Daisuke Satow}
\email{d-sato@ruby.scphys.kyoto-u.ac.jp}

\author{Yoshimasa Hidaka}
\email{hidaka@ruby.scphys.kyoto-u.ac.jp}

\author{Teiji Kunihiro}
\email{kunihiro@ruby.scphys.kyoto-u.ac.jp} 
\affiliation{Department of Physics, Faculty of Science, Kyoto University, Kitashirakawa Oiwakecho, Sakyo-ku, Kyoto 606-8502, Japan}

\begin{abstract}
 We investigate spectral properties of a fermion
 coupled with a massive gauge boson with a  mass  $m$ at finite temperature ($T$) 
in the perturbation theory.
The massive gauge boson is introduced as a  $U(1)$ gauge boson in 
the Stueckelberg formalism with a gauge parameter $\alpha$.
We find that the fermion spectral function has a 
three-peak structure for  $T \sim m$ irrespective of the choice
of the gauge parameter, while it tends to have one faint peak at the origin and two peaks corresponding to the normal fermion and anti-plasmino excitations
familiar in QED in the hard thermal loop approximation
for $T \gg m$. 
We show that our formalism successfully describe
the fermion spectral function in the whole $T$ region with the correct high-$T$ limit except for the faint peak at the origin,
although some care is needed for  choice of 
 the gauge parameter for $T \gg m$.
We clarify that for $T \sim m$, 
the fermion pole is almost independent of the gauge parameter 
in the  one-loop order,
while for $T \gg m$, 
the one-loop analysis is valid only for $\alpha \ll 1/g$ where $g$ is the fermion-boson
coupling constant, implying that the one-loop analysis can not be valid
for large gauge parameters as in the unitary gauge.
\end{abstract}

\date{\today}

\maketitle

\section{Introduction}
\label{sec:introduction}

It is well known that  
for extremely high temperature ($T$) where the  hard thermal loop (HTL)
approximation in QED and QCD \cite{frenkel-taylor,braaten-pisarski,weldon:1982aq,weldon} is valid,
a fermion (quark) coupled with thermally excited gauge fields (gluons) 
make collective excitations, i.e., the normal fermion (particle)  and the 
anti-plasmino excitation
with distinct peaks in the fermion spectral function \cite{weldon};
this feature obtained in the HTL approximation is also known 
to be gauge invariant in the sense 
that the fermion self-energy at one-loop order does not depend on gauge \cite{braaten-pisarski}.
As for lower $T$ region,
a possible change in the spectral properties of the quark in association with 
chiral transition in QCD was investigated \cite{kitazawa-NJL}, using the Nambu-Jona-Lasinio 
model \cite{NJL}, and  it is  shown that the coupling with the chiral 
soft modes \cite{kunihiro} make the quark spectral function have
 distinct three peaks  near but above the critical temperature of chiral transition.
The appearance of such a novel spectral function at $T \sim m$ 
 was later confirmed \cite{kitazawa}
for
 a massless fermion coupled with an elementary massive boson with a mass $m$,
 irrespective of the type of the massive boson.
The mechanism for realizing the three-peak structure in the spectral function was 
also elucidated \cite{kitazawa} in terms of the Landau damping owing to the collisions of the fermion with
thermally excited bosons\footnote{
This  feature that the three-peak structure arises at $T \sim m$ is not altered 
even for a massive fermion with a mass $m_f$
as long as $m_f$ is not too large compared with $m$ \cite{mitsutani}.
}.

Then one may  naturally ask a question if the fermion spectral function at those lower $T$ 
would smoothly connect with that at extremely high $T$, i.e. the HTL result in QED/QCD:
If it is not the case, it means that we do not have a unified understanding of the 
fermion spectral properties in the whole $T$ region.
Partly to answer this  question, 
we investigate  spectral properties of a fermion coupled with
a massive vector boson introduced  as a $U(1)$ gauge boson
in the (generalized) Stueckelberg 
formalism with a gauge parameter $\alpha$ \cite{stueckelberg-review,stueckelberg},
and  carefully examine their  possible gauge dependence at $T\not=0$, 
at the one-loop order  as in \cite{kitazawa}.
Here the spectral properties include 
the number of the fermion poles,  the pole position in the complex energy plane 
and the spectral function in the momentum-energy plane. We are also interested in how
the quasi-particle nature of the fermion is realized or destroyed by the coupling 
with a massive boson at finite $T$.

We find that the present formalism gives a valid description of the fermion coupled with a
massive vector boson for
 the  whole temperature ($T$) region at one-loop order in a unified way; thereby
we reveal the characteristics of the fermion spectral properties depending on  
the distinct $T$ regions, i.e.
(I)\,$T \ll m$,\, (II)\,$T \sim m$ and (III)\,$T \gg m$. 
Especially, we shall show that 
 the fermion spectral function certainly 
tends to have a three-peak structure for $T\sim m$ in the small momentum region
 with  supports in the positive, zero and negative
energy regions.

The investigation of the possible gauge dependence 
turns out to be involved owing to the appearance of a novel mass scale $\sqrt{\alpha} m$,
 inherent in the present formalism, as well as the boson mass $m$ and temperature $T$.
One should remark here that the Proca formalism adopted in \cite{kitazawa}
 is not adequate for this purpose, 
because this formalism corresponds to a special gauge with $\alpha \rightarrow \infty$ 
(unitary gauge), and
does not lead to the proper high-$T$ limit, or $m/T \rightarrow 0$, which should be 
the HTL approximation in QED at one-loop level \cite{weldon}.
This is the reason why we have adopted the Stueckelberg formalism to
describe the massive vector boson.
We remark that although the pole position  is gauge-independent in the exact 
calculation \cite{rebhan},
a gauge-dependence of the fermion pole may appear in the perturbation
theory at finite $T$ in general.
Since the Proca formalism corresponding to the limit $\alpha \rightarrow \infty$ 
leads to a wrong high-$T$ limit,
 there should exist
an adequate gauge-parameter region in which the results in the perturbation theory
hardly show gauge dependence: Indeed, we show that this is the case in the present work.


This paper is organized as follows.
In Sec.~\ref{sec:formalism},
 we formulate the $U(1)$ gauge theory in which the gauge boson acquires finite mass.  
We perform a calculation of a fermion self-energy at finite temperature.
In Sec.~\ref{sec:results}, the numerical results of the fermion 
spectral properties are shown.
In Sec.~\ref{sec:gauge}, we discuss the gauge dependence of fermion pole 
appearing when $T \gg m$ in an analytic way.
Section~\ref{sec:summary} is devoted to a summary and concluding remarks.
In Appendix~\ref{app:higgs}, we  briefly describe
how the abelian Higgs model is reduced to the massive  gauge theory 
in the Stueckelberg formalism.
In Appendix~\ref{app:self-energy}, we present detailed calculational procedures for the fermion self-energy 
in our model.
Appendix~\ref{app:Blimit} is devoted to making an order estimate 
of some terms appearing in the text.

\section{U(1) gauge theory with massive gauge boson}
\label{sec:formalism}

In this section, we formulate the $U(1)$ gauge theory with a massive gauge boson,
 and introduce a propagator and a spectral function at finite temperature 
in the imaginary time formalism~\cite{lebellac,kapusta}.
We perform a calculation of the self-energy of a fermion coupled with a massive vector boson at one-loop order.

\subsection{General formalism}
First, we introduce a $U(1)$ gauge theory with a massive gauge boson.
The gauge boson acquires a mass by the Higgs mechanism, keeping the gauge symmetry.
The gauge theory is one way to construct a renormalizable quantum field theory with a massive vector boson.
We employ the Stueckelberg formalism~\cite{stueckelberg,stueckelberg-review} proposed long ago, 
which is equivalent to the abelian Higgs model with a constant absolute value of 
the Higgs field~\cite{stueckelberg-review,higgs-stueckelberg}.
This correspondence is reviewed in Appendix~\ref{app:higgs}.
Then our Lagrangian reads
\begin{equation} 
\begin{split}
\label{eq:formalism-lagrangian}
{\cal L}&=-\frac{1}{4}F_{\mu\nu}F^{\mu\nu} + \frac{1}{2}m^2 \left(A_\mu-\frac{\partial_\mu B}{m}\right)\left(A^\mu -\frac{\partial^\mu B}{m}\right) 
+\overline{\psi}(i(\partial_\mu -ig A_\mu)\gamma^\mu)\psi +{\cal L}_{\text {GF}},
\end{split}
\end{equation}
where $A_\mu$, $B$ and $\psi$ are  a massive vector, a scalar and a fermion field, respectively.
The scalar field $B$ is called the Stueckelberg field,
 which corresponds to the phase of the Higgs field in the abelian Higgs model.
 $F_{\mu \nu}=\partial_\mu A_\nu-\partial_\nu A_\mu$ is a field strength, 
$g$ the coupling constant, $m$ the vector boson mass, and $\alpha$ is a gauge parameter.
 ${\cal L}_{\text {GF}}$ is the gauge fixing term defined by
\begin{equation}
{\cal L}_{\text {GF}}\equiv-\frac{1}{2\alpha}(\partial_{\mu} A^\mu +\alpha m B)^2.
\end{equation}

We work with the Minkowski metric, $g_{\mu \nu}=\mathrm{diag}\,(1,\, -1,\, -1,\, -1)$.
We shall deal with a massless  fermion assuming that the mass is neglected,
which should be valid at high temperatures. 
Our Lagrangian is invariant under the gauge transformation except for the gauge fixing term, ${\cal L}_{\text {GF}}$:
\begin{align}
\psi(x)& \rightarrow e^{ig\Lambda(x)}\psi(x),\\
 A_\mu (x)& \rightarrow A_\mu (x)+\partial_{\mu} \Lambda (x),\\
B(x)& \rightarrow B(x) +m \Lambda(x).
\end{align}
There are no interaction between the Stueckelberg field and the fermion field,
 and we chose the gauge fixing term so that the interaction term between the vector field and the Stueckelberg field vanishes.
We can drop the Stueckelberg field  as long as a correlation function is concerned,
while it can not be when the thermodynamic potential is considered,
where it is important to take into account the correct degrees of freedom.

The propagator of the free massive vector boson is now given by
\begin{equation}
D_{\mu\nu}(p)=\frac{-1}{p^2-m^2}
\left(
g_{\mu\nu}-\frac{p_\mu p_\nu}{p^2-m^2\alpha}(1-\alpha)
\right).
\end{equation}
In the $\alpha \rightarrow \infty$ limit, the propagator tends to
\begin{equation}
D_{\mu\nu}(p)\rightarrow \frac{-1}{p^2-m^2}
\left(
g_{\mu\nu}-\frac{p_\mu p_\nu}{m^2}
\right),
\end{equation}
which is  the massive vector-boson propagator in the Proca formalism\footnote{
Here we note that the propagator 
in the Proca formalism does not vanish but rather 
approaches a constant value in the $p\rightarrow \infty$ limit, in contrast to
that in the Stueckelberg formalism.
This causes the non-renormalizability and leads to a
 bad behavior at high temperature~\cite{proca-problem,kitazawa,dolan-jackiw}.
 }.

The fermion propagator $G(p)$ in the imaginary time formalism~\cite{kapusta,lebellac} is expressed with the self-energy $\Sigma(p)$ as
\begin{equation}
G(p)=\frac{1}{\Slash{p}-\Sigma(p)}, 
\end{equation}
where  $p^0=i\omega_m=i(2m+1)\pi T$ is the Matsubara frequency for  fermion.
Note that $G(p)$ and $\Sigma(p)$ are $4\times 4$ matrices with the spinor indices.
The retarded fermion propagator is given by an analytic continuation, $i\omega_m\to \omega+i\epsilon$:
\begin{equation}
G^{\text R}\pw =  G(\vect{p},\omega+i\epsilon)=
\frac{1}{\omega \gamma^0-{\vect{p}\cdot \vgamma}-\Sigma^{\text R}\pw},
\end{equation}
where the retarded self-energy is given by 
\begin{equation}
\Sigma^{\text R}\pw = \Sigma(\vect{p},\omega+i\epsilon).
\label{eq:retarded-Sigma}
\end{equation}
Introducing the projection operator 
on the (anti-)particle sector 
$\Lambda_{\pm}(\vect{k})=(1\pm\gamma^0{ \vgamma \cdot \hat{\vect{k}}})/2$,
 we can decompose the retarded propagator and  self-energy into the respective sector as follows:
\begin{align} 
G^{\text R}\pw&=G_+\pw\Lambda_+(\vect{p})\gamma^0+G_-\pw\Lambda_-(\vect{p})\gamma^0,\\
\label{eq:helicity-selfenergy}
\Sigma^{\text R}\pw&=\Sigma_+\pw\Lambda_+(\vect{p})\gamma^0+\Sigma_-\pw\Lambda_-(\vect{p})\gamma^0,
\end{align}
with $\Sigma_{\pm}\pw={\text {Tr}}\,(\Sigma^{\text R}\pw \Lambda_{\pm}(\vp)\gamma^0)/2$.

In the particle sector, the pole $\omega_\vp=\omega(\vect{p})$ satisfies the following equation:
\begin{equation}
G^{-1}_+(\vp,\omega_\vp)=\omega_\vp-|\vect{p}|-\Sigma_+(\vp,\omega_\vp)=0.
\label{eq:dispersionRelation}
\end{equation}
From the analyticity of the retarded propagator,
the pole is located on the real axis or the lower half-plane of  complex $\omega$.
If the imaginary part of the pole is small,
the pole is well described in terms of a quasi-particle picture, where
the real part of the pole corresponds to the energy while 
the imaginary part to the decay width of the quasi-particle.
If the imaginary part is large,
then it would be meaningless to consider excitations in terms
of any particle picture.
\begin{figure}[t]
\begin{center}
\includegraphics[width=0.35\textwidth]{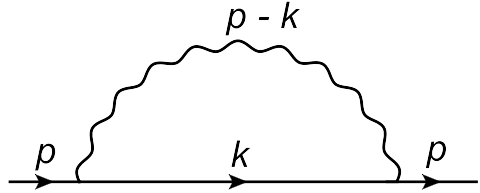}
\caption{The diagram which contributes to the fermion self-energy at one-loop order.
The solid line represents the fermion and the wavy line represents the massive vector boson.}
\label{fig:selfenergy}
\end{center}
\end{figure}
It is known that the self-energy at zero momentum has the following symmetry,
\begin{align}
\label{eq:formalism-symmetry-realsigma}
\re\,\Sigma_+(\vect{0},-\omega)&=-\re\,\Sigma_+(\vect{0},\omega),\\
\label{eq:formalism-symmetry-imagsigma}
\im\,\Sigma_+(\vect{0},-\omega)&=\im\,\Sigma_+(\vect{0},\omega),
\end{align}
which implies that 
if there exists a fermion pole at $z=\omega-i\Gamma$ at zero momentum, 
there is also a pole at $-\omega-i\Gamma$ at zero momentum.

Once the self-energy  $\Sigma_\pm\pw$ is obtained,
the spectral function of the (anti-)particle sector is expressed 
  as
\begin{align}
\label{eq:formalism-spectral}
\begin{split}
\rho_\pm\pw&=-\frac{1}{\pi}\,\im\,G_\pm\pw
=-\frac{1}{\pi}\frac{\im\,\Sigma_\pm\pw}{(\omega\mp|\vect{p}|-\re\,\Sigma_\pm\pw)^2+\im\,\Sigma^2_\pm\pw} .
\end{split}
\end{align}
When the peak is narrow enough, 
the position of the peak is given by $\omega_\vp=|\vect{p}|+\re\,\Sigma_+(\vp,\omega_\vp)$
 and the width of the peak is given by $-\,\im\,\Sigma_+(\vp,\omega_\vp)/(2\omega_\vp)$.

\subsection{Calculation at one-loop order}

Now let us evaluate the self-energy $\Sigma(p)$ at one-loop order;
the corresponding diagram is shown in Fig.~\ref{fig:selfenergy}.
$\Sigma(p)$ is expressed as
\begin{align}
\begin{split}
\Sigma(\vect{p},i\omega_m)&=-g^2T\sum_{n}\int \frac{d^3\vect{k}}{(2\pi)^3}\gamma^{\mu}G_0(\vect{k},i\omega_n)\gamma^{\nu} 
D_{\mu\nu}(\vect{p}-\vect{k},i\omega_m-i\omega_n) \\
&=g^2T\sum_{n}\int \frac{d^3\vect{k}}{(2\pi)^3}\gamma^{\mu}\frac{\Slash{k}}{k^2}\gamma^{\nu}
\frac{1}{l^2-m^2}\left(
g_{\mu\nu}-(1-\alpha)\frac{l_\mu l_\nu}{l^2-m^2\alpha}
\right),
\end{split}
\end{align}
where $G_0$ is the propagator of the free fermion, $l\equiv p-k$, and $k^0=i\omega_n=i(2n+1)T$.
Some manipulations lead to
\begin{equation}
\begin{split}
\Sigma(\vect{p},i\omega_m)&=-2g^2\gamma^\mu \tilde{B}_\mu\imagm 
+\frac{g^2}{m^2}\Bigl[\Slash{p}\Bigl(p^2\left(\tilde{B}\imagalpham-\tilde{B}\imagm\right)\\
&\quad-m^2\left(\alpha \tilde{B}\imagalpham-\tilde{B}\imagm\right)\Bigr) 
-p^2\gamma^\mu\left(\tilde{B}_\mu\imagalpham-\tilde{B}_\mu\imagm\right)\Bigr].
\end{split}
\label{eq:stueckelberg-selfenergy-tensor}
\end{equation}
The retarded self-energy $\Sigma^{\text R}(\mathbf{p}, \omega)$ in the one-loop approximation
is given by the analytic continuation $i\omega_m\rightarrow \omega+i\epsilon$
from $\Sigma(\vect{p},i\omega_m)$.
Here, we have introduced the following loop functions:
\begin{align} 
\tilde{B}\imagm\equiv&\;T\sum_{n}\int \frac{d^3\vect{k}}{(2\pi)^3}
\frac{1}{(k-p)^2-m^2}\frac{1}{k^2}, \\
\tilde{B}^\mu\imagm\equiv&\;T\sum_{n}\int 
\frac{d^3\vect{k}}{(2\pi)^3}\frac{k^\mu}{(k-p)^2-m^2}\frac{1}{k^2}.
\end{align}
We see that there are two kinds of mass in Eq.~(\ref{eq:stueckelberg-selfenergy-tensor}),
 $m$ and $\sqrt{\alpha}m$,
 the latter of which is unphysical because it depends on the gauge parameter.
However,
the existence of such an unphysical mass causes two different high temperature limit
 as will be shown in Sec.~\ref{sec:gauge}.
We will also show that $\Sigma(p)$ approaches the fermion self-energy in QED if we take 
the massless limit $m\rightarrow 0$, which is not the case in the Proca formalism.
It should be noted here that 
Eq.~(\ref{eq:stueckelberg-selfenergy-tensor}) shows that there is a special value 
of $\alpha$:
 when $\alpha=1$, the terms containing the unphysical mass are all cancelled out and 
only the first term remains, i.e.,
$\Sigma(\vect{p},i\omega_m)_{\alpha=1}=-2g^2\gamma_\mu \tilde{B}^\mu\imagm$.

The self-energy in the (anti-)particle sector in the one-loop approximation 
is given by $\Sigma_{\pm}\pw={\text {Tr}}\,(\Sigma^{\text R}\pw \Lambda_{\pm}(\vp)\gamma^0)/2$.
Then, as is derived in Appendix~\ref{app:self-energy},
 we have for the imaginary part of $\Sigma_+(\vect{p},\omega)$,
\begin{equation}
\label{eq:stueckelberg-imsigma-exact}
\begin{split}
\im\,\Sigma_+(\vect{p},\omega)&=-\frac{g^2}{32\pi|\vect{p}|^2m^2}\int^{E'^{-}_{f}}_{E'^{+}_{f}}dE_f(f(E_f)+n(E_f-\omega))
 [(-p^2+m^2\alpha)(|\vect{p}|-\omega)^2+2p^2E_f(\omega-|\vect{p}|)]\\
&\quad+\frac{g^2}{32\pi|\vect{p}|^2m^2}\theta(-p^2)[p^2(\omega-|\vect{p}|)\pi^2 T^2 +
\omega(\omega-|\vect{p}|)(-\omega p^2-(\omega-|\vect{p}|)(-p^2+m^2\alpha))]\\
&\quad+\frac{g^2}{32\pi|\vect{p}|^2m^2}\int^{E^{-}_{f}}_{E^{+}_{f}}dE_f(f(E_f)+n(E_f-\omega))[(-p^2+m^2)((|\vect{p}|-\omega)^2-2m^2) +2(p^2-2m^2)E_f(\omega-|\vect{p}|)]\\
&\quad-\frac{g^2}{32\pi|\vect{p}|^2m^2}\theta(-p^2)[(p^2-2m^2)(\omega-|\vect{p}|)\pi^2 T^2+\omega[2m^4-p^2(|\vect{p}|(\omega-|\vect{p}|)+m^2)]],
\end{split}
\end{equation}
where $p^2=\omega^2-{ |\vect{p}|}^2$, 
$E^{\pm}_{f}=(\omega^2-|\vect{p}|^2-m^2)/(2(\omega\pm|\vect{p}|))$ and ${E'}^{\pm}_{f}=(\omega^2-|\vect{p}|^2-\alpha m^2)/(2(\omega\pm|\vect{p}|))$.

The real part $\re\, \Sigma_+(\mathbf{p}, \omega)$ 
may be obtained using the dispersion relation from the imaginary part.
Especially, the finite temperature part of the real part of the self-energy,
$\Sigma_+(\vect{p},\omega)_{T\neq 0}\equiv\Sigma_+(\vect{p},\omega)-\Sigma_+(\vect{p},\omega)_{T=0}$, 
is expressed as
\begin{equation}
\label{eq:formalism-dispersion}
\re\,\Sigma_+(\vect{p},\omega)_{T\neq 0}=-\frac{1}{\pi}{\text P}\int^\infty_{-\infty}d\omega'\frac{\im\,\Sigma_+(\vect{p},\omega')_{T\neq 0}}{\omega-\omega'}.
\end{equation}
Here P denotes the principal value.
The zero temperature part of Re$\Sigma_+(\vect{p},\omega)$ is
 not determined by Eq.~(\ref{eq:formalism-dispersion}) because it has ultraviolet divergence.
We make renormalization using twice-subtracted dispersion relation, which reads
\begin{equation} \begin{split}
\re\,\Sigma_+(\vect{p},\omega)_{T=0}&=c_0+c_1(\omega-|\vect{p}|)
+\frac{(\omega-|\vect{p}|)^2}{\pi}{\text P}\int^\infty_{-\infty}dz\frac{\im\,\Sigma_+(\vect{p},z)_{T=0}}{(z-|\vect{p}|)^2(z-\omega)}.
 \end{split}\end{equation}
We impose the on-shell renormalization condition, $\Sigma_+(\vect{p},\omega=|\vect{p}|)=0$
 and $\partial\Sigma_+(\vect{p},\omega)/\partial\omega|_{\omega=|\vect{p}|}=0$, to determine $c_0$ and $c_1$.
The vacuum part of $\im\,\Sigma_+(\vect{p},\omega)$ is obtained by taking the $T\rightarrow 0$ limit of Eq.~(\ref{eq:stueckelberg-imsigma-exact});
\begin{equation}
\begin{split}
\im\,\Sigma_+(\vect{p},\omega)_{T=0}&=\frac{g^2}{32\pi m^2}\frac{\sgn(\omega)}{p^2}(\omega-|\vect{p}|)
\left[\theta(p^2-\alpha m^2)(p^2-\alpha m^2)^2
-\theta(p^2-m^2)\frac{(p^2+2m^2)(p^2-m^2)^2}{p^2}\right].
\end{split}
\end{equation}
Thus we arrive at
\begin{equation}
\begin{split}
\re\,\Sigma_+(\vect{p},\omega)_{T=0}=&\frac{g^2p^2}{32\pi^2m^2}(\omega-|\vect{p}|)
\Bigl[\frac{2m^4}{p^4}-(2+\alpha)\frac{m^2}{p^2}+\ln\left|\frac{p^2-m^2}{p^2-m^2\alpha}\right|
 +\alpha\frac{m^2}{p^2}\left(-2+\frac{m^2\alpha}{p^2}\right)\ln\left|\frac{\alpha(p^2-m^2)}{p^2-m^2\alpha}\right| \\
&+\frac{m^2}{p^2}\left(-2\alpha+(3+\alpha^2)\frac{m^2}{p^2}-2\frac{m^4}{p^4}\right)\ln\left|\frac{m^2}{p^2-m^2}\right|\Bigr]. 
\end{split}
\end{equation}

As mentioned before,
our theory based on the Stueckelberg formalism approaches QED
at high enough temperature where the masses are negligible in comparison with $T$.
Let us see this.
For $T \gg gT \gg m,\sqrt{\alpha}m$, the imaginary and real part
of  the self-energy are reduced to
\begin{align}
\label{eq:stueckelberg-hight-htl-im}
\begin{split}
\im\,\Sigma_+(\vect{p},\omega)_{T\rightarrow \infty}&\simeq\frac{g^2}{32\pi|\vect{p}|^2m^2}\theta(-p^2)[p^2(\omega-|\vect{p}|)\pi^2 T^2 
-(p^2-2m^2)(\omega-|\vect{p}|)\pi^2T^2] \\
&=\frac{g^2\theta(-p^2)}{16|\vp|^2}\pi T^2(\omega-|\vect{p}|),
\end{split}
 \\
\label{eq:stueckelberg-hight-htl}
\re\,\Sigma_+(\vect{p},\omega)_{T\rightarrow \infty}&\simeq\frac{g^2T^2}{16|\vp|^2}
\left(2|\vect{p}|+(|\vect{p}|-\omega){\ln}\left|\frac{\omega+|\vect{p}|}{\omega-|\vect{p}|}\right|\right),
\end{align}
respectively.
Here, we have retained only  the terms 
which are proportional to $T^2$ in Eq.~(\ref{eq:stueckelberg-imsigma-exact}).
These Eqs.~(\ref{eq:stueckelberg-hight-htl-im}) and 
(\ref{eq:stueckelberg-hight-htl}) coincide exactly 
with the well-known results in the HTL approximation in 
QED~\cite{frenkel-taylor,braaten-pisarski,weldon:1982aq,weldon}.
There is a caveat in the above manipulation, which has been taken for granted
in the usual derivation of the HTL approximation in the gauge theory:
The ignored terms may  become comparable to terms which are proportional 
to $T^2$ in some gauges and hence the above naive power-counting turns out to be invalid.
We will analyze this possibility in Sec.~\ref{sec:gauge}.

\section{Numerical Results}
\label{sec:results}
In this section, we show  numerical results of the fermion spectral function
 and the fermion poles at various temperatures.
In the following, the coupling constant is fixed to a small value, $g=0.5$,
 so that the analysis based on the one-loop calculation can be valid:
Except when  the coupling constant dependence of the pole is analyzed,
the coupling constant will be fixed to $g=0.5$.
On the other hand,
the gauge parameter, $\alpha$, will be 
varied freely in order to see the gauge-dependence of the spectral properties of the 
fermion calculated at one-loop level.
\subsection{Low temperature ($T \ll m$) }
\label{ssc:results-low-T}

In this subsection, we show  numerical results at a so low temperature that 
$T$ dependence of the results is hardly seen, 
which may check our analytical and numerical 
calculations.

\begin{figure}[!t]
\begin{center}
\includegraphics[width=0.5\textwidth]{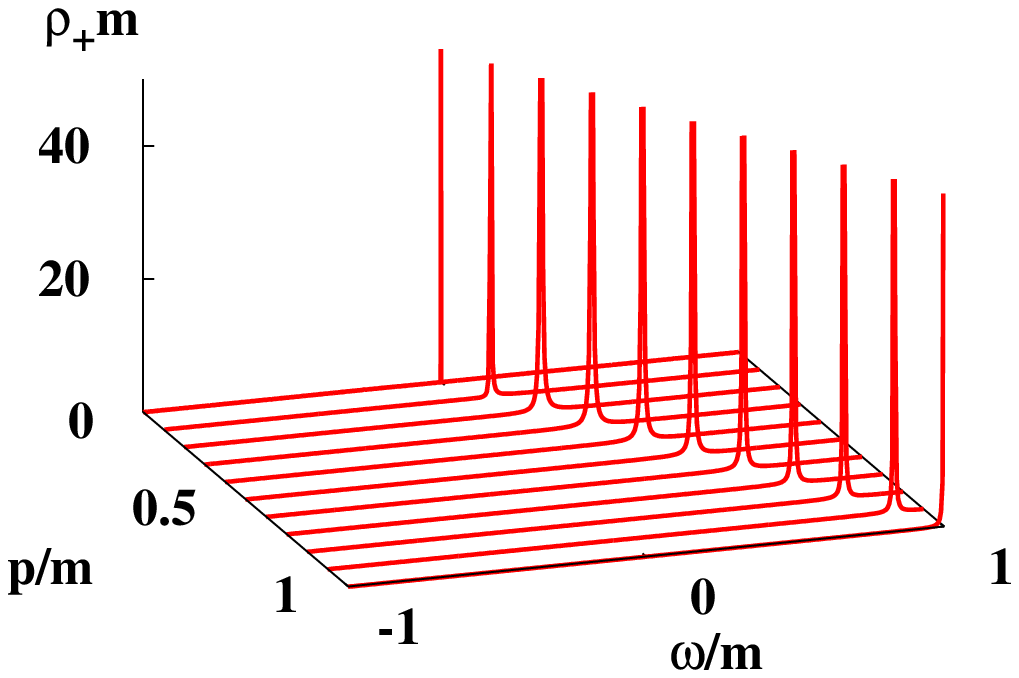}
\caption{The fermion spectral function in the particle sector $\rho_+$
 as a function of energy $\omega$ and momentum $p$ for $T=0.4\,m,~ g=0.5,~ \alpha=1$.}
\label{fig:vac}
\end{center}
\end{figure}

 Figure~\ref{fig:vac} shows the fermion spectral function in the particle sector
(with a positive particle number)
at $T=0.4\,m$ for $\alpha=1$. 
There appears a very narrow peak near $\omega=|\vect{p}|$, 
which is very reminiscent of zero temperature case.
This is natural for $T \ll m$, because the thermal effect is exponentially suppressed 
by the Boltzmann factor $\sim \exp(-m/T)$, and hence the breaking of Lorentz symmetry is small.
This small breaking of Lorentz symmetry implies that the particle pole is 
almost on-shell value at $T=0$, i.e., $\omega=|\vect{p}|$,
 and hence the gauge dependence of the pole hardly appears.

\begin{figure*}[!t]
\begin{center}
\includegraphics[width=\textwidth]{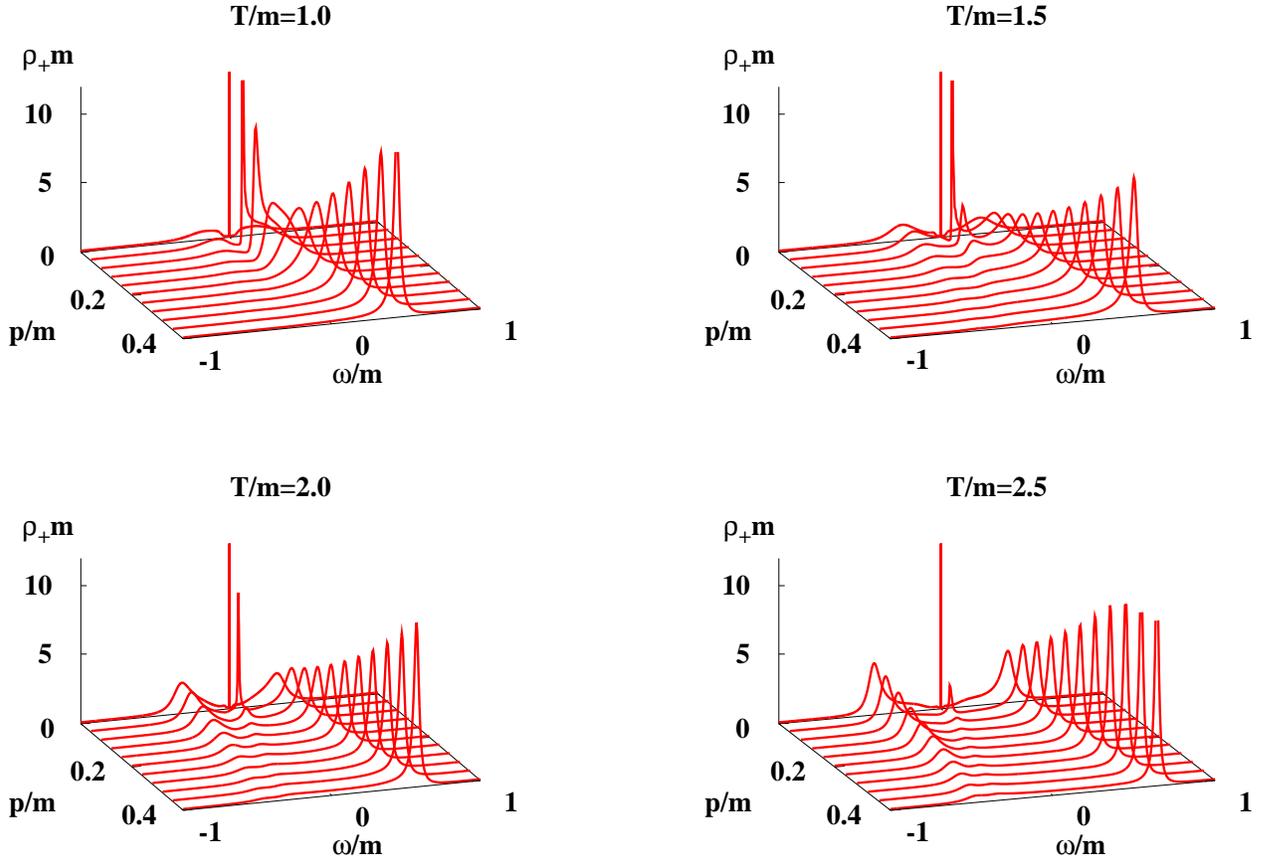}
\caption{The fermion spectral function in the particle
 sector $\rho_+$ as a function of energy $\omega$ and momentum 
$p$ at $g=0.5$,~ $\alpha=1$ at various temperatures.}
\label{fig:3peak-cont}
\end{center}
\end{figure*}

\begin{figure}[!h]
\begin{center}
\includegraphics[width=0.5\textwidth]{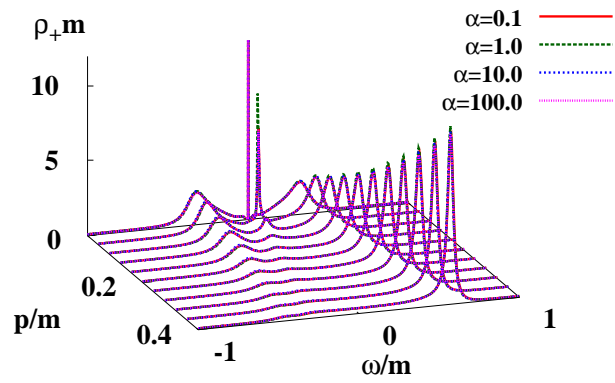}
\caption{The spectral fermion function in the particle sector $\rho_+$ 
as a function of energy $\omega$ and momentum $p$ at $T=2.0\,m, ~g=0.5$.}
\label{fig:0.5-2}
\end{center}
\end{figure}

\begin{figure*}[t]
\begin{center}
\includegraphics[width=\textwidth]{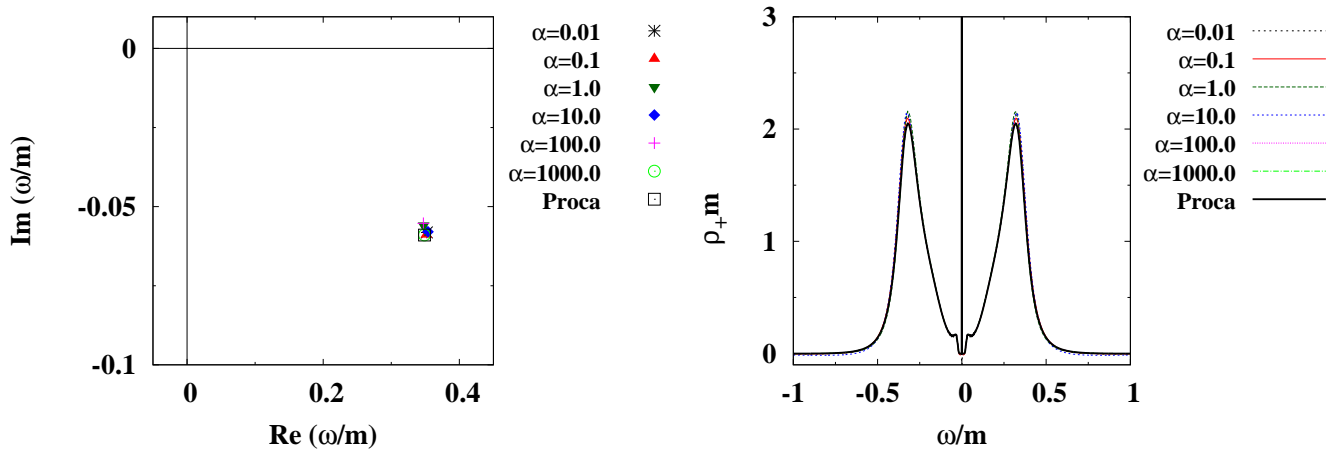}
\caption{The left panel shows the gauge dependence of the fermion pole with a positive
particle number in the positive energy region for $T=2.0\,m$, $g=0.5$ at zero momentum:
 the horizontal and vertical axes denote the real and imaginary part of the energy, respectively.
The right panel shows the fermion spectral function in the particle sector $\rho_+$
 as a function of energy $\omega$ for the same $T$, $g$ as the left panel.}
\label{fig:3peak}
\end{center}
\end{figure*}

\begin{figure*}[!t]
\begin{center}
\includegraphics[width=140mm]{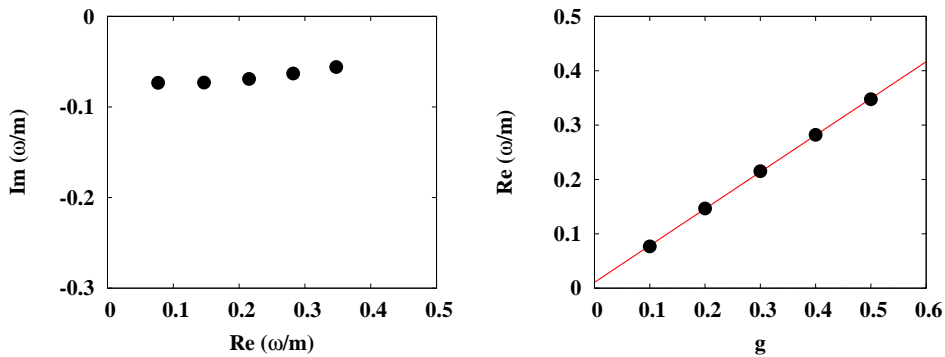}
\caption{The left panel shows the coupling constant dependence of the fermion pole with
a particle number 
in the positive energy region for $T/m=2.0$ and $\alpha=1.0$, at zero momentum: 
the horizontal and the vertical axis denote the real and imaginary part of the energy,
respectively.
The dots from left to right correspond to $g=0.1$, $0.2$, $0.3$, $0.4$, $0.5$, 
respectively.
The right panel shows the coupling constant dependence of the real part of the fermion(particle) 
pole in the positive energy region at zero momentum for the same $T/m$ and $\alpha$.
The solid line is the fitted linear function.}
\label{fig:pole-g}
\end{center}
\end{figure*}

\begin{figure}[t]
\begin{center}
\includegraphics[width=0.45\textwidth]{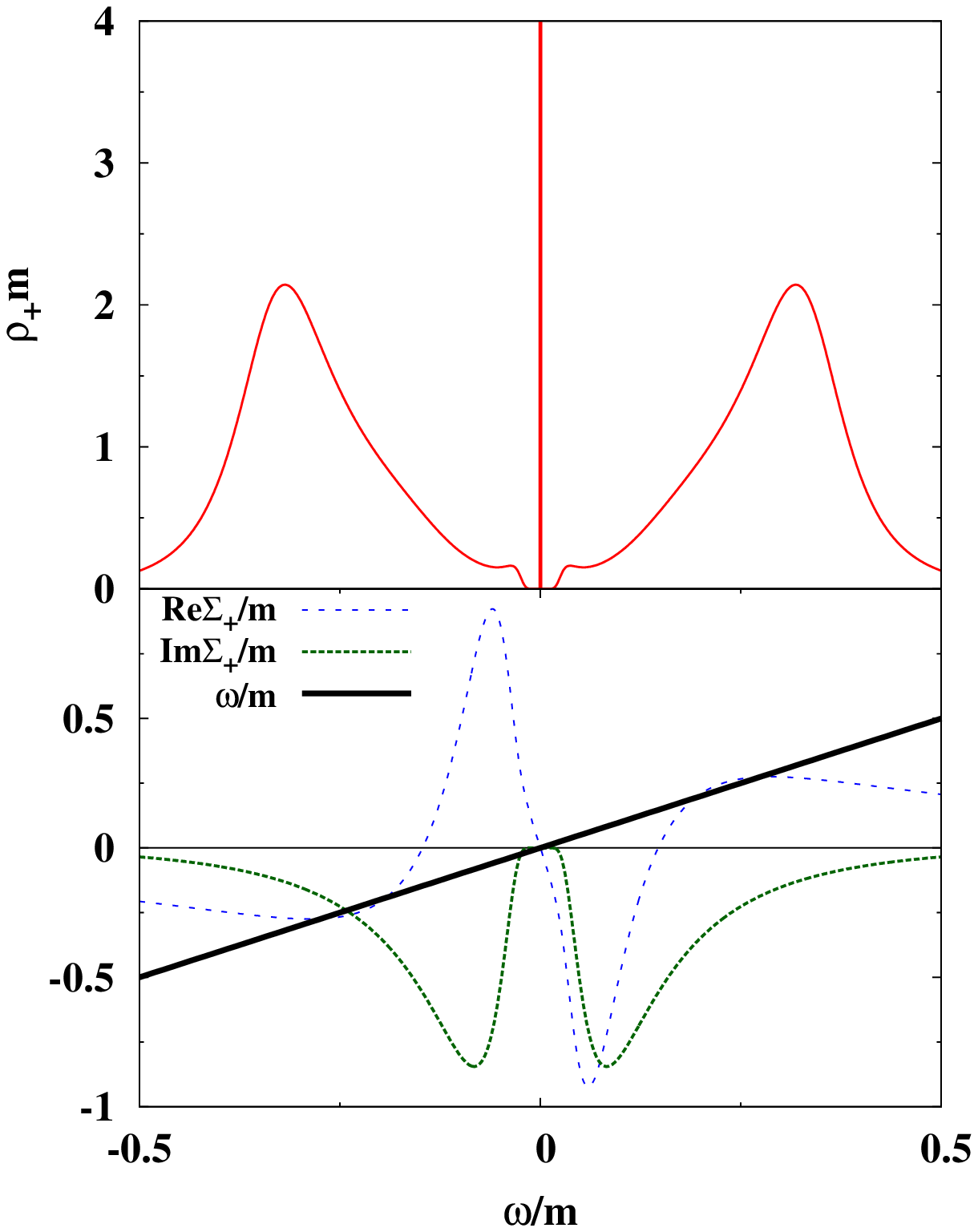}
\caption{The upper figure shows the fermion spectral function in the particle sector 
$\rho_+$ as a function of energy $\omega$ for $g=0.5, T=2.0m$, and $\alpha=1$.
The lower figure shows the real part and the imaginary part of the fermion self-energy
 as functions of energy $\omega$ for the same $g$, $T$, and $\alpha$.  
The dotted line denotes $\omega$.
The intersection points between this line and
 ${\text {Re}}\Sigma_+(\vect{0},\omega)$ are the solutions of the dispersion relation.}
\label{fig:3peak-analysis}
\end{center}
\end{figure}
\subsection{Intermediate temperature ($T \sim m$)}
\label{ssc:results-intermediate-T}

We plot the spectral function of the particle sector at
 $T/m=1.0$, $1.5$, $2.0$, $2.5$ for $\alpha=1$ in Fig.~\ref{fig:3peak-cont}.
We can see that the spectral function at these temperatures have  structures 
qualitatively different from that at low temperature:
Even at $T/m=1.0$, we see a split of a peak around the origin seen in Fig.~\ref{fig:vac}
 into two peaks  with a small bump in the
{\em negative} energy region, which is reminiscent of the anti-plasmino peak known in QED/QCD at high
$T$. These features are enhanced as $T$ is  raised, and we see 
 a clear three-peak structure around $|\vp|=0$  at $T/m=2.0$ with a prominent peak and a clear bump in the
positive and negative energy region, respectively.
We also see that the peak near the origin is attenuated as $T$ is further raised
up to $2.5\, m$.

Since it is known that
the details of the shape of the spectral function may be gauge-dependent in general,
let us  see how  the three-peak structure depends on the gauge parameter.
 Figure~\ref{fig:0.5-2} shows the gauge-parameter dependence of
 the fermion spectral function in the particle sector at $T=2.0\,m$;
the gauge parameter is varied as $\alpha=0.1$, $1$, $10$ and  $100$.
One might find only single curve of the spectral function in the figure, 
although this figure actually shows four curves of it with
different $\alpha$; thus it clearly tells us 
that the shape of the spectral function at $T=2\,m$ with a three-peak structure 
is virtually independent of the gauge parameter.

The virtual gauge-independence of the shape of the spectral function implies that 
the pole of the propagator is also the case.
We  show the gauge (in)dependence of the pole in the positive energy region 
at $|\vp|=0$ with a particle number
in the left panel of Fig.~\ref{fig:3peak}, which shows that the pole position is almost independent of the choice of the gauge parameter, as anticipated:
Note that the gauge parameter is varied in a wider range than in Fig.~\ref{fig:0.5-2},
i.e.,  $\alpha=0.01$, $0.1$, $1$, $10$, $100$, $1000$.
A remark is in order here: 
The pole in the negative energy region at $|\vp|=0$ has the same properties as that in the positive
energy region, as is assured by
Eqs.~(\ref{eq:formalism-symmetry-realsigma}) and (\ref{eq:formalism-symmetry-imagsigma}).

Such a gauge-independence of the poles necessarily reflects in that of the spectral function.
The right panel of Fig.~\ref{fig:3peak} shows the fermion spectral function at
 zero momentum for the  wide range of $\alpha$ up to $1000$, 
together with that obtained in  the Proca formalism\footnote{
The rapid decrease of the spectral function in $|\omega|\leq0.1$ is caused by the exponential damping 
of $\im\,\Sigma_+(\vect{0},\omega)$ in that region.}.
From this figure, we confirm  that the spectral function at zero momentum is 
virtually gauge-independent for the wide range of $\alpha$.

We also note that the position and the width of the peaks coincide with
the real and imaginary part of the poles, respectively, which is due to the fact
that the imaginary part of the poles is small in comparison with the real part, 
as seen in the left panel of Fig.~\ref{fig:3peak}.
Thus the shape of 
the spectral function with a three-peak structure necessarily 
gets to have almost no gauge-dependence.

We show the coupling constant dependence of the fermion pole at zero momentum 
in Fig.~\ref{fig:pole-g} for $T/m=2.0$.
The real part is almost proportional to $g$, like that in QED in the HTL approximation.
The coupling constant dependence of the imaginary part is not large.

What is the mechanism for realizing the three-peak structure of the fermion spectral function?
Figure~\ref{fig:3peak-analysis} shows the real and imaginary part of the self-energy
for $T/m=2.0$ and $\alpha=1.0$ at $|\vp|=0$, together with the corresponding spectral function.
A detailed analysis of the imaginary part tells us that the peaks of the imaginary part
correspond to a Landau damping of the fermion by a scattering with thermally excited bosons.
Since these features of the fermion self-energy is very similar to that shown in 
\cite{kitazawa}, the mechanism for realizing the three-peak structure found in our 
formalism  is understood to be the same as discussed in \cite{kitazawa}.


\subsection{High temperature ($T \gg m$) }
\label{ssc:results-high-T}

In this subsection,
 we show numerical results in the high temperature ($T \gg m$) region,
where the mass of the vector boson (and the fermion) can be neglected in comparison with
$T$, i.e., $m/T\rightarrow 0$; this means that $T$ itself may not be infinitely large.

\begin{figure*}[!t]
\begin{center}
\includegraphics[width=\textwidth]{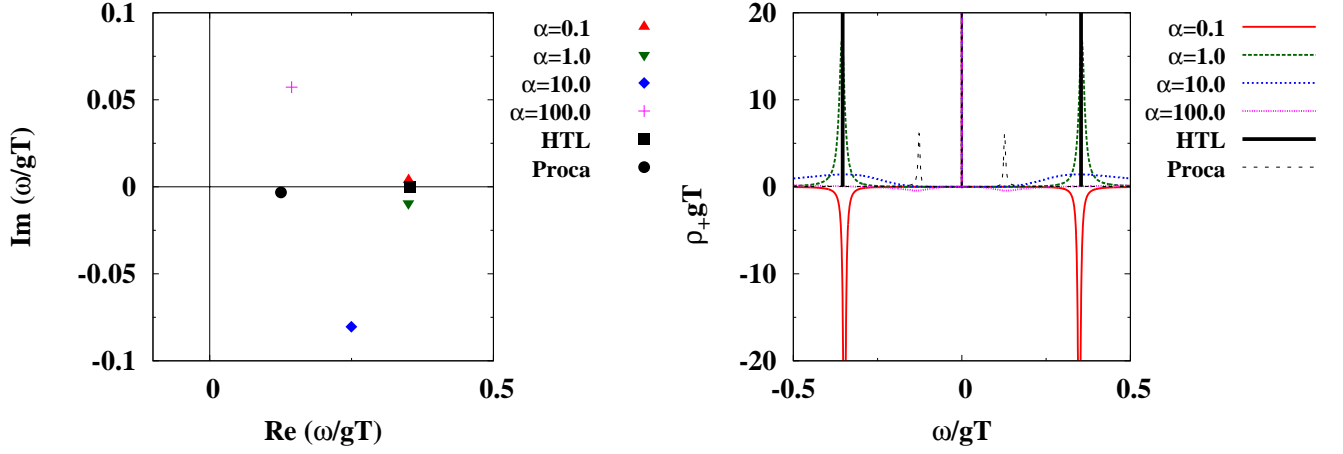}
\caption{The left panel shows the gauge dependence of the fermion pole at $p=0$ 
in the positive energy region at $T=40.0m$ for $g=0.5$ :
 The horizontal and the vertical axis denote the real and imaginary part of energy.
The right panel shows the fermion spectral function in the particle sector $\rho_+$ 
as a function of energy $\omega$  for the same $T$ and  $g$.}
\label{fig:pole-10}
\end{center}
\end{figure*}

We show the fermion pole in the positive energy region in the left panel of 
Fig.~\ref{fig:pole-10} at $T=40.0\,m$ and for $\alpha=0.1,1,10$ and $100$.
The pole in the Proca formalism and that in the HTL approximation in QED are also shown.
We see that the gauge dependence of the fermion pole is no longer negligible. 
Since the exact pole position in the complex energy plane should be
 gauge-independent~\cite{rebhan}, 
the above result suggests that the one-loop analysis is no longer valid 
in this high-$T$ region in contrast to
the $T \ll m$ and $T \sim m$ regions, at least in some gauge.
We will present a
detailed discussion on 
how the gauge dependence arises at high $T$ region in Sec.~\ref{sec:gauge}.

One should notice that the pole for $\alpha=0.1$ is located in the upper energy plane, 
which could be problematic because it 
 implies a loss of the analyticity of the retarded propagator
and also negativeness of the  spectral function, 
 as  seen from Eq.~(\ref{eq:formalism-spectral}).

\begin{figure}[!t]
\begin{center}
\includegraphics[width=0.5\textwidth]{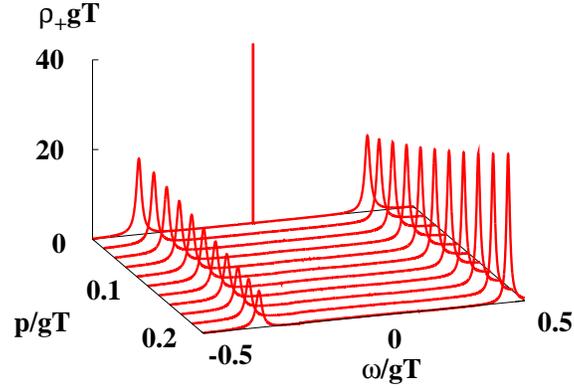}
\caption{The fermion spectral function in the particle sector $\rho_+$ as a function of energy 
$\omega$ and momentum $p$ at $T=40.0m$ with $g=0.5$ and $\alpha=1$.}
\label{fig:2peak}
\end{center}
\end{figure}

For $\alpha \sim 1$, there appear clear  two peaks 
 in the spectral function in a robust way,
as shown in Fig.~\ref{fig:2peak}; 
the two peaks are found to tend to the normal fermion(particle) and the anti-plasmino 
of QED in  the HTL approximation~\cite{frenkel-taylor,braaten-pisarski,weldon:1982aq,weldon}, respectively;
see Sec.~\ref{sec:gauge}.

There persists the other peak at the origin in the energy-momentum space.
One can confirm that its residue is of the order of $m^4/(g^2T^4)$, which is very small if we consider the $gT\gg m$ case, by making power counting.
Such a peak at the origin  was also obtained in \cite{kitazawa}, though in 
the Proca formalism. 
One should also remark that
such a peak at the zero energy is not obtained in QED
 in the HTL approximation, 
in which the vector boson mass is set to zero from the beginning, 
in contrast to the present case\footnote{
The absence of a peak at the vanishing energy in QED with
 the HTL approximation is easily understood  as follows:
In the HTL approximation of QED, 
Re$\Sigma_+(\vect{0}, \omega)$ behaves as $1/\omega$
 in the $\omega\rightarrow 0$ limit.
Thus at $\omega=0$, the pole condition Eq.~(\ref{eq:dispersionRelation}) will not be 
satisfied and hence there can not exist a pole at the origin.}.
It should be intriguing to explore whether 
this peak at the origin extends to a finite-$|\vect{p}|$ region, and
hence the three-peak structure of the fermion spectral function persists
even in such a high-$T$ region, i.e., for $T \gg m$.
In fact, this is a challenging problem in quantum field theory at finite temperature,
because a sensible analysis of such an infrared region 
requires a systematic method to remove the so called pinch singularities \cite{pinchSingularity}. 
This task is beyond the scope of the present work, and 
we leave such an  analysis as a future work \cite{persistency}.

Our numerical calculation has shown 
that one can have virtually gauge-independent results even in the one-loop analysis
if the gauge parameter is in the region $\alpha \sim 1$. 
We shall argue that 
the perturbative expansion should be valid for $\alpha \ll 1/g$ in Sec.~\ref{sec:gauge}.
It means that 
the spectral function of a fermion coupled with a massive vector boson 
as calculated in the Stueckelberg formalism nicely approaches 
that in QED in the HTL approximation at high $T$ irrespective of the choice
of the gauge parameter $\alpha$, if the order of $\alpha$ is confined to 
$\alpha\ll 1/g$. 
This is actually already suggested 
by the asymptotic form Eq.~(\ref{eq:stueckelberg-hight-htl}) 
for $T \gg m, \sqrt{\alpha}\,m$.

\section{Analysis of Gauge dependence of the Pole at high temperature ($T \gg m$)}
\label{sec:gauge}

Our numerical calculation has shown that 
the pole position of the fermion propagator 
is virtually independent of the gauge parameter $\alpha$
for the cases of $T \ll m$ and $T \sim m$:
The former case is simply because
 the thermal contribution due to a boson with a mass $m$
is greatly suppressed by a Boltzmann factor $\exp(-m/T)\ll 1$
when $T \ll m$.
By contrast, for $T\gg m$, the numerical results in Sec.~\ref{ssc:results-high-T} 
show that the pole of the fermion propagator has a large gauge dependence for large $\alpha$.
In this section, we discuss the gauge dependence of the pole 
of the fermion propagator at weak coupling at high temperature.
In particular, we focus on the region $gT\gg m$.
In this region, one expect that the mass of the vector boson can be neglected, and
thus the self-energy approach that in the HTL approximation of 
QED \cite{frenkel-taylor,braaten-pisarski,weldon:1982aq,weldon},
in which the fermion has the pole of order $gT$. 
Therefore, we analyze the pole of the fermion propagator by assuming $\omega\sim gT$.
Here we introduce a small dimensionless parameter, 
\begin{equation}
\mOvergT \equiv\left(\frac{m}{gT}\right)^2 \ll 1\,.
\end{equation}
Thus we have two small dimensionless parameters, $g$ and $\lambda$,
which are treated  as independent parameters, so that
the self-energy is expanded by combined powers of $g$ and $\lambda$.
If the the power of $g$ and $\mOvergT$ are both positive, 
the high temperature limit will be well defined
and smoothly connected to that of QED.
However, as will be shown below, an inverse power of $\mOvergT$ appears
at one loop level when the gauge parameter is large, and hence
 the high temperature limit becomes inevitably different from that of QED.

In the following analysis, we put $|\vp|=0$ for simplicity. 
The pole position obtained in the perturbation theory 
generally depends on the gauge parameter as well as
 $g$, $\lambda$ and $T$
due to the truncation of the perturbative expansion.
We parametrize the pole $\omega_{\text{pole}}$ of the fermion propagator as 
\begin{equation}
\omega_{\text{pole}} = gT F(g,\lambda,\alpha) ,
\end{equation}
where $F(g,\lambda,\alpha)$ is a function of order one,
 and depends on the gauge parameter $\alpha$.
If the limit,
\begin{equation} 
F_0 \equiv \lim_{g\to0}F(g,\lambda,\alpha), 
\end{equation}
is independent of $\alpha$, then the pole is independent of the gauge parameter 
at the  order $gT$.
Thus one sees that the  gauge dependent part may be  defined by
\begin{equation}
\delta \omega_{\text{pole}}(g,\lambda,\alpha) \equiv  \omega_{\text{pole}}
(g,\lambda,\alpha) - \omega_{\text{pole}}^0(\lambda),
\label{eq:delOmegaPole}
\end{equation}
where $\omega_{\text{pole}}^0(\lambda) \equiv gT F_0(\lambda)$.
For a reference, we recall that  $F_0(\lambda=0)=1/(2\sqrt{2})$
in the case of QED~\cite{weldon}.
When the inequality, 
\begin{equation}
\omega_{\text{pole}}^0(\lambda) \gg \delta \omega_{\text{pole}}(g,\lambda,\alpha),
\label{eq:poleCondition}
\end{equation}
is satisfied, the gauge dependence can be neglected.
In reality with a finite $g$, the region of the gauge parameter
 satisfying Eq.~(\ref{eq:poleCondition}) will be limited.
We shall call the region that the gauge parameter satisfies Eq.~(\ref{eq:poleCondition})
 as {\it an adequate gauge parameter region}.
The purpose of this section is to find the adequate gauge parameter region.

\begin{figure}[t]
\begin{center}
\includegraphics[width=0.45\textwidth]{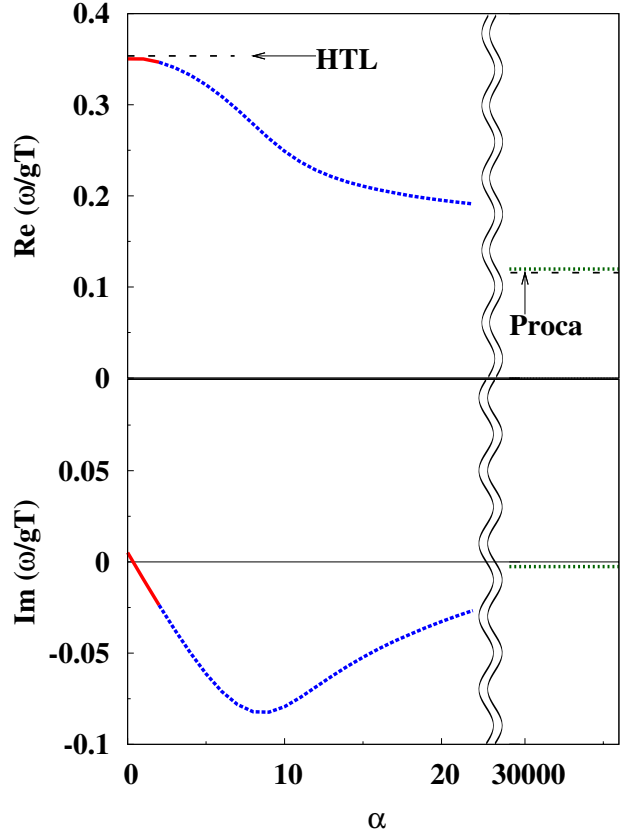}
\caption{The upper panel shows the gauge dependence of the real part of the pole in the
 positive energy region at zero momentum at $T=40.0m$, for $g=0.5$. 
 The horizontal and the vertical axes denote the gauge parameter $\alpha$ and the real
 part of the pole, respectively.
 The data whose $\alpha$ is smaller than the upper limit of adequate gauge parameter, $1/g=2.0$, are denoted by the solid line and the data whose $\alpha$ is larger than $1/g$ by the dotted line. 
 The data which satisfy $\alpha > 1/(g^2\mOvergT)=1600$, which are expected to approach the value in the Proca formalism, are plotted with another dotted line.
The upper and lower dashed lines show the thermal mass in the HTL approximation $gT/(2\sqrt{2})$ 
and  that in the Proca formalism $\sqrt{6}m/\sqrt{1+48\lambda}\simeq2.31\,m$, respectively.
The lower panel shows the gauge dependence of the imaginary part of the pole for the same $T$
 and $g$ as in the upper panel.}
\label{fig:pole-cont}
\end{center}
\end{figure}

Let us first show a numerical result of the real and the imaginary part of the pole at 
$T=40.0\,m$ as functions of $\alpha$ in Fig.~\ref{fig:pole-cont}:
For a large $\alpha$ ($1 \ll \alpha $), the $\alpha$ dependence of the real part 
of the pole is large, and especially for very large $\alpha$, say $\alpha \sim 3\times 10^4$, the magnitude of it is no longer of $O(gT)$, but is of a smaller order, $O(m)$, as will be  shown later.
The imaginary part of the pole for $\alpha\ll 1$ is positive and apparently problematic
because it means that the analyticity of the retarded propagator is lost and the fermion spectral
 function will become negative.
As we shall show later, however, the absolute value of the 
imaginary part is of $O(g^2T)$ and should be considered 
together with  
higher order contributions. So the negative imaginary part  with a small absolute value
can be ignored in this order of the coupling.

Now we shall show that such an order estimate of the pole can be done analytically.
We start with an analysis of the self-energy, under the condition that $\omega\sim gT$,
by decomposing the self-energy (\ref{eq:stueckelberg-selfenergy-tensor}) 
to seven parts,
\begin{equation}
\Sigma_+({\mathbf 0},\omega)\equiv \omega\,\sig{}= \omega\sum_{n=1}^{7}\sig{n},
\end{equation}
where we have introduced the following dimensionless functions:
\begin{align}
\label{eq:results-tensor}
\one&=-\frac{2g^2}{\omega}\tilde{B}^0(\vz, \omega;m)  ,\\
\label{eq:results-tensor-2}
\two&=+\frac{g^2\omega^2}{m^2}\tilde{B}(\vz, \omega;\sqrt{\alpha}m) ,\\
\label{eq:results-tensor-3}
\three&=-\frac{g^2\omega^2}{m^2}\tilde{B}(\vz, \omega;m) ,\\
\label{eq:results-tensor-4}
\four&=-g^2\alpha\tilde{B}(\vz, \omega;\sqrt{\alpha}m) ,\\
\label{eq:results-tensor-5}
\five&=+g^2\tilde{B}(\vz, \omega;m) ,\\
\label{eq:results-tensor-6}
\six&=-\frac{g^2\omega}{m^2}\tilde{B}^0(\vz, \omega;\sqrt{\alpha} m) ,\\
\label{eq:results-tensor-7}
\seven&=+\frac{g^2\omega}{m^2}\tilde{B}^0(\vz, \omega; m). 
\end{align}
Here $\tilde{B}(\vz, \omega;m)$ and $\tilde{B}^0(\vz, \omega;m)$ are obtained 
by performing the analytic continuation ($i\omega_m\rightarrow \omega+i\epsilon$) 
to $\tilde{B}(\vz,i\omega_m;m)$ and $\tilde{B}^0(\vz,i\omega_m;m)$.
From Eq.~(\ref{eq:dispersionRelation}), one sees that the poles satisfy
the condition
\begin{equation}
C(\omega_{\text{pole}})=1 .
\label{eq:poleCondition2}
\end{equation}

\begin{figure}
\begin{center}
\includegraphics[width=0.5\textwidth]{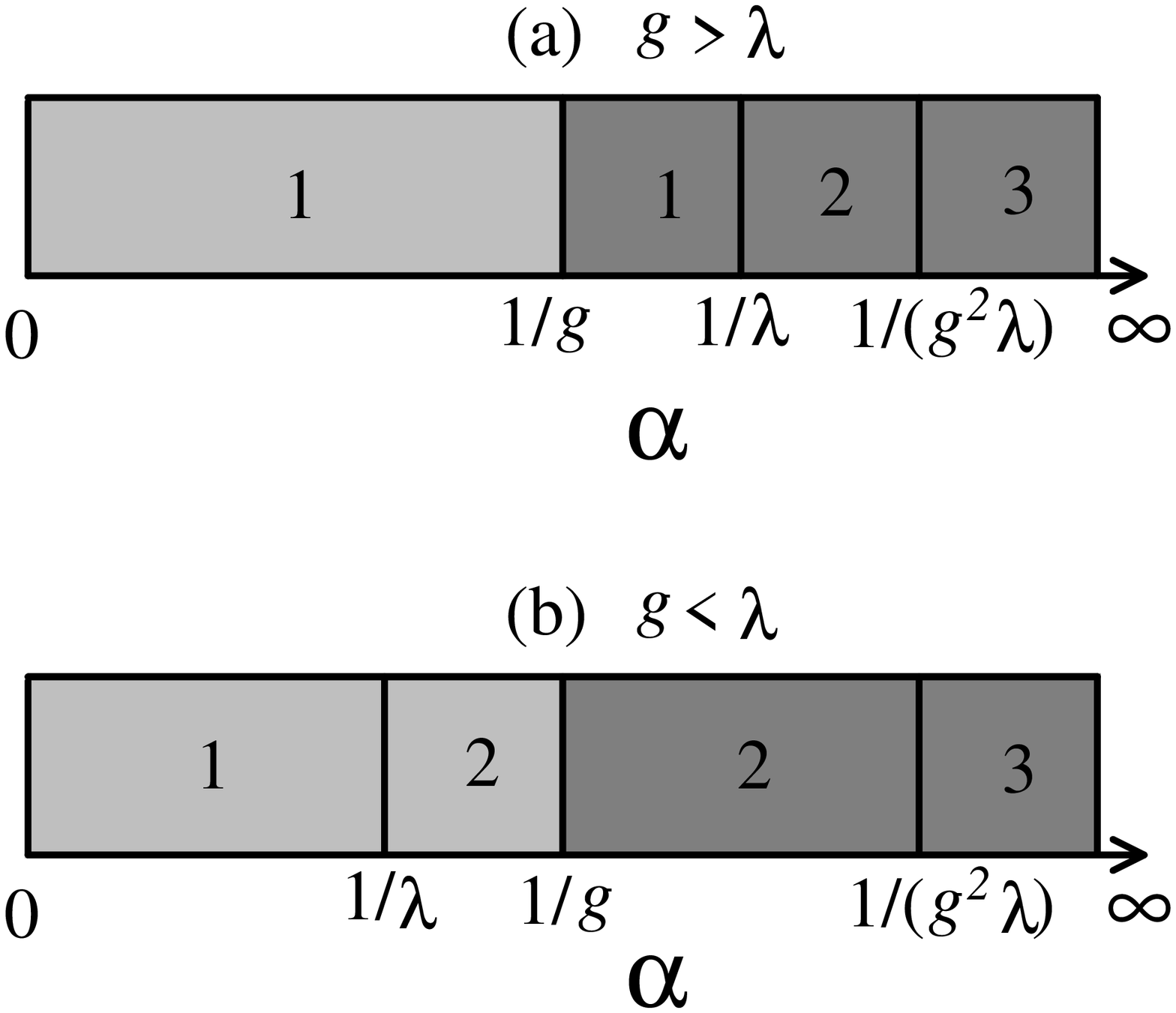}
\caption{The classification of the gauge parameter regions 
depending on the relative magnitude of 
$\mOvergT =m^2/(gT)^2$ to $g$ at high temperature, i.e., for $gT\gg m$ or $\lambda \ll 1$.
Each number in the figures corresponds to the case with same number in the text.
The light gray regions are the adequate gauge parameter regions for which
Eq.~(\ref{eq:poleCondition}) is satisfied, 
whereas the dark gray region are not.}
\label{fig:perturbation-analysis}
\end{center}
\end{figure}
We first note that if the following equation has a root, 
\begin{equation}
\lim_{g\to0}C(gTF) = 1 ,
\end{equation}
there is a pole $\omega_{\text{pole}}=gTF$ of order $gT$.
Furthermore,
if the $F$ happens to be independent of $\alpha$, the pole
is gauge independent.
Therefore,
 let us take the following function
as a measure of the gauge dependence, instead of Eq.~(\ref{eq:delOmegaPole}):
\begin{equation}
\delta C(g,\lambda,\alpha) \equiv C(\omega_{\text{pole}}^0)-1 .
\end{equation}
Then a criterion for the adequate gauge parameter region may be
given by 
\begin{equation}
\delta C(g,\lambda,\alpha) \ll 1.
\end{equation}

We now make an order estimate of the seven terms defined in Eqs.~(\ref{eq:results-tensor}) $\sim$ (\ref{eq:results-tensor-7}). 
One finds that this task is reduced to that of  $\tilde{B}(\vz,\omega;m)$ and $\tilde{B}^0(\vz,\omega;m)$.
The following relations are shown in Appendix~\ref{app:Blimit}:
\begin{itemize}
\item When $T \gg \omega \gg M$, 
\begin{equation}
\tilde{B}(\vz,\omega;M)\sim \frac{T}{\omega},\quad \frac{\tilde{B}^0 (\vz,\omega;M)}
{\omega}\sim \frac{T^2}{\omega^2}.
\label{eq:hight}
\end{equation}
\item When $T \gg M \gg \omega$, 
\begin{align}
\begin{split}
\tilde{B}(\vz,\omega;M)&\sim \left\{
\begin{array}{ll}
\frac{T}{\omega} \quad &\text{for} \quad M^2 \ll \omega T\\
\frac{T^2}{M^2}  \quad &\text{for} \quad M^2 \gg \omega T
\end{array}
\right.,\\
\frac{\tilde{B}^0 (\vz,\omega;M)}{\omega}&\sim \left\{
\begin{array}{ll}
\frac{T^2}{\omega^2}  \quad &\text{for} \quad M^2 \ll \omega T\\
\frac{T^4}{M^4}  \quad &\text{for} \quad M^2 \gg \omega T
\end{array}
\right..
\label{eq:middlet}
\end{split}
\end{align}
\item When $M \gg T \gg \omega$,
\begin{equation}
\tilde{B}(\vz,\omega;M)\sim \frac{T^2}{M^2} ,\quad  \frac{\tilde{B}^0 (\vz,\omega;M)}{\omega} \sim \frac{T^4}{M^4}.
\label{eq:lowt}
\end{equation}
\end{itemize}
Here, $M$ denotes $m$ or $\sqrt{\alpha}m$, and the vacuum parts have been dropped.
Using these relations, we will find the adequate gauge parameter region in the following 
subsections, and
the result is summarized in Fig.~\ref{fig:perturbation-analysis},
 which shows that the adequate gauge parameter is $\alpha\ll 1/g$.

$\one$, $\three$, $\five$, and $\seven$, which do not depend on the gauge parameter,
are estimated to be
\begin{align}
\one \simeq g^2 \frac{T^2}{8\omega^2}\sim 1,\quad 
\three\sim g\left(\frac{1}{\mOvergT}\right),\quad
\five\sim g,\quad
\seven\sim  \left(\frac{1}{\mOvergT}\right).
\end{align}
We remark that $\one$ coincides with the fermion self-energy
 in the HTL approximation in QED, as it should be.

We note that $\three$ and $\seven$ are of the order of  an inverse power of $\lambda$, which would make
it impossible to take the massless limit.
These `dangerous' terms are found to be nicely canceled out with
 other $\sig{i}$'s when $\alpha$ is not so large, whereas
for  large $\alpha$, the cancellation does not happen, and the condition $\delta C\ll 1$ can not
 be satisfied, as will be shown below.
\subsection{$1 \gg \alpha\mOvergT$ {\text {(case 1)}}}

When $1 \gg \alpha\mOvergT$,
the mass scale $m$ and $\sqrt{\alpha}m$ are negligible in comparison with  $gT$,
and then the self-energy coincides with 
that in HTL approximation in QED in this approximation,
because the non-leading terms can be neglected as we have seen in Sec.~\ref{sec:formalism}.

On account of the order estimate Eq.~(\ref{eq:hight}),
we obtain
\begin{equation}
\two\sim g\left(\frac{1 }{\mOvergT}\right),\quad
\four\sim g\alpha,\quad
\six\sim \left(\frac{1}{\mOvergT}\right).
\label{eq:tgtalphamm}
\end{equation}
The leading terms of $\two$ and $\six$  cancel out\footnote{As seen from 
Eqs.~(\ref{eq:app-T-gg-omega-gg-m-B-Ef<T*-2}) and (\ref{eq:app-T-gg-omega-gg-m-B0-Ef<T*}), 
$\tilde{B}(p;m)$ and $\tilde{B}^0(p;m)$ yield terms which are proportional to $m$, so $\two$,
 $\three$, $\six$, and $\seven$ seems to yield terms which are proportional to $m^{-1}$ and 
the massless limit ($m/T\rightarrow 0$) can not be taken.
Actually, from Eqs.~(\ref{eq:app-T-gg-omega-gg-m-B-Ef<T*-2}) and
 (\ref{eq:app-T-gg-omega-gg-m-B0-Ef<T*}), we see that the terms discussed above cancel out. 
Therefore we can take the massless limit and the fermion self-energy approaches that in QED.} with  $\three$ and $\seven$,
respectively, and 
the terms of the order $g\alpha$ and $g$ remain\footnote{These terms come from the imaginary part of $\tilde{B}(\vz,\omega;m)$, $\tilde{B}^0(\vz,\omega;m)$, $\tilde{B}(\vz,\omega;\sqrt{\alpha}m)$, and $\tilde{B}^0(\vz,\omega;\sqrt{\alpha}m)$, which can be confirmed by retaining the next-to-leading term in Eqs.~(\ref{eq:app-T-gg-omega-gg-m-B-im-first}) and (\ref{eq:app-T-gg-omega-gg-m-B0-im}).
The contribution from the real parts are much smaller than these terms.
}
.
Thus one sees that the gauge dependent part is of the order of $g\alpha$, which means that 
the pole is gauge independent in practice, which also can be confirmed from Fig.~\ref{fig:pole-cont},  provided that 
the inequality  $\alpha \ll 1/g$ is satisfied;
in this case, the adequate gauge parameter region is 
the region of $\alpha$ which satisfies the above inequality.

Let us see that the imaginary part of the self-energy is of order $g^2T$ in the Landau gauge ($\alpha=0$).
For $\alpha=0$, the imaginary part of the self-energy is evaluated to be
\begin{equation}
\begin{split}
{\text {Im}}\Sigma_{+}(\mathbf{0},\omega)&=\frac{g^2}{64\pi m^2}
\left[\omega^3\left(\tanh\frac{\omega^2}{4T\omega}+\coth\frac{\omega^2}{4T\omega}\right)
-\frac{(\omega^2+2m^2)(\omega^2-m^2)^2}{\omega^3}
\left(\coth\frac{\omega^2+m^2}{4T\omega}+
\tanh\frac{\omega^2-m^2}{4T\omega}\right)\right]\\
&\simeq\frac{1}{16\pi}g^2T,
\end{split}
\end{equation}
where the first line is obtained by substituting $\vp=0$ and $\alpha=0$ in Eq.~(\ref{eq:app-imsigma}).
In the second line the inequality $T\gg \omega\gg m$ was used.
Although this is positive and apparently breaks the analyticity,
this order of the coupling should not be determined in the one-loop order, because 
 the two-loop diagrams contain contributions of order $g^2T$.
In fact, the analyticity problem 
can be cured by taking into account the two-loop
 diagrams. 

\subsection{$1/g^2 \gg \alpha \mOvergT \gg 1$ {\text {(case 2)}} }
\subsubsection{$\alpha \ll 1/(g{\mOvergT})$ (case 2-a)}
On account of the order estimate  for $\two$, $\four$, and $\six$
given in Eq.~(\ref{eq:middlet}), we obtain
\begin{equation}
\two\sim\, g\left(\frac{1}{\mOvergT}\right),\quad
\four\sim\, g\alpha,\quad
\six\sim\, \left(\frac{1}{\mOvergT}\right),
\end{equation}
which is the same as Eq.~(\ref{eq:tgtalphamm}). 
The leading term of $\six$  cancels out
with that of $\seven$ and 
the remaining terms  are of the order 
of $g\alpha$\footnote{This term comes from Im$\tilde{B}^0(\vz,\omega;\sqrt{\alpha}m)$, which can be confirmed from Eq.~(\ref{eq:app-T-gg-m-gg-omega-B0-im}).
The real parts are negligible compared with this term.}
, while $\two$ and $\three$ do not.
In the present case, however, $\four$ is larger than $\two$ and $\three$,
and hence $\four$ dominates the gauge-dependent terms,
which should be made small.
This smallness is guaranteed when $\alpha \ll 1/g$,
which defines  the adequate gauge parameter region;
this coincides with that  in the case 1.
Notice that the inequality  $g < \mOvergT$ must be assumed 
in this case, otherwise the inequality $\alpha<1/g$ can not be satisfied. 

\subsubsection{$\alpha \gg 1/(g\mOvergT)$ (case 2-b)}

Again, on account of the order estimate in Eq.~(\ref{eq:middlet}),
 $\two$, $\four$, and $\six$ are estimated to be
\begin{equation}
\two\sim \frac{1}{\alpha}\left(\frac{1}{\mOvergT}\right)^2,~
\four\sim \left(\frac{1}{\mOvergT}\right) ,~
\six\sim \frac{1}{\alpha^2g^2\mOvergT^3}.
\end{equation}
Though $\four$ and $\seven$ have the same order of magnitude, they do not cancel out, which can be confirmed from Eqs.~(\ref{eq:app-T-gg-omega-gg-m-B0}) and (\ref{eq:app-T-gg-m-gg-omega-B}).
Therefore the largest contribution is $\four\sim1/\lambda\gg1$;
the present region for the gauge parameter ($\alpha \gg 1/(g\mOvergT)$)
is not an adequate gauge parameter region.

\subsection{$\alpha \mOvergT \gg 1/g^2 $ {\text {(case 3)}} }

Here we treat the case where the gauge parameter is far larger than
$1/\lambda g^2$; this case includes the unitary gauge ($\alpha \rightarrow \infty$).
Owing to the order estimate Eq.~(\ref{eq:lowt}) for this case,
 $\two$, $\four$, and $\six$ are estimated to be
\begin{equation}
\two\sim \frac{1}{\alpha}\left(\frac{1}{\mOvergT}\right)^2,\quad
\four\sim \left(\frac{1}{\mOvergT}\right),\quad
\six\sim \frac{1}{g^2\alpha^2\mOvergT^3}.
\end{equation}
Again, although $\four$ and $\seven$ have the same order of magnitude, 
they do not actually cancel out on account of the difference in the coefficients, which can be seen from Eqs.~(\ref{eq:app-T-gg-omega-gg-m-B0}) and (\ref{eq:app-m-gg-T-gg-omega-B}).
In this case, the largest contribution is $\four\sim1/\lambda\gg1$;
therefore this region is not an adequate gauge parameter region.
This also suggests that the order of the pole in the unitary gauge is not $gT$. 

So let us now discuss the pole in the unitary gauge.
First we assume the pole is of order $m$ instead of $gT$.
Using Eqs.~(\ref{eq:lowt}), (\ref{eq:app-T-gg-omega-sim-m-B}), and
 (\ref{eq:app-T-gg-omega-sim-m-B0}), we have the following order estimates:
\begin{align}
\one\sim\, \frac{1}{\mOvergT},\quad
\two\sim\, -\frac{1}{\alpha\mOvergT},\quad
\three\sim\, \frac{g}{\sqrt{\mOvergT}},\quad
\four\sim\,\frac{1}{\mOvergT},\quad
\five\sim\, \frac{g}{\sqrt{\mOvergT}},\quad
\six\sim\, \left(\frac{1}{\alpha g \mOvergT}\right)^2,\quad
\seven\sim\,-\frac{1}{\mOvergT}.
\end{align}
For $\alpha\to\infty$, $\two$ and $\six$ vanish.
 Furthermore $\three$ and $\five$ can be also neglected, because $g \sqrt{\lambda}=m/T \ll 1$.
The remaining parts $\one$, $\four$ and $\seven$
 are estimated more precisely with the use of 
 Eqs.~(\ref{eq:app-T-gg-omega-gg-m-B0}) and (\ref{eq:app-m-gg-T-gg-omega-B}), as follows:
\begin{equation}
\one\simeq\frac{1}{8\lambda}\frac{m^2}{\omega^2}, \quad
\four\simeq\frac{1}{24\lambda},\quad
\seven\simeq-\frac{1}{16\lambda}.
\end{equation}
Collecting these terms, we reach at
\begin{equation}
\sig{} \simeq \frac{1}{48\lambda} \left( \frac{6m^2}{\omega^2} - 1 \right).
\end{equation}
This is precisely the same as the result obtained  in the
 Proca formalism  in the high temperature limit~\cite{kitazawa};
the thermal mass in this case reads 
\begin{equation}
\label{eq:fermion-pole-proca}
\omega=\frac{\sqrt{6}\,m}{\sqrt{1+48\lambda}} \xrightarrow{\lambda\rightarrow 0}\sqrt{6}m,
\end{equation}
which can be seen from Fig.~\ref{fig:pole-cont} also.
We have confirmed numerically that the spectral function also approaches that
 in the Proca formalism  as $\alpha\to\infty$.

\subsection{Brief Summary}

Let us summarize and discuss the results obtained so far in the preceding subsections for 
the gauge-parameter dependence of the fermion propagator at high temperature.
The results are summarized in Fig.~\ref{fig:perturbation-analysis}.
We have found an adequate gauge parameter region as $\alpha \ll 1/g$ in which possible gauge
dependence is of higher order of the couplings and hence can be neglected.
We remark that this parameter restriction should also apply to QED.

In fact, the electron self-energy in the HTL approximation in QED
 at next-to-leading level in the one-loop order reads~\cite{wang,Mottola}
\begin{align}
\Sigma^{\text R}(\vp={\mathbf 0},\omega)\simeq&\frac{e^2T^2}{8\omega}\gamma^0+
\frac{e^2}{8\pi^2}\gamma^0\left(\omega\ln\frac{T}{\omega}-i\pi T\right)
+(1-\alpha)\frac{e^2}{8\pi^2}\gamma^0\left(-\omega\ln\frac{T}{\omega}+\frac{3\pi iT}{2}\right).
\label{eq:HTLQED}
\end{align}
If it were that $\omega\sim eT$ and $\alpha\sim 1/e$, the third term would be the same order as the
 leading term, which implies that the gauge-independence is badly broken.
Conversely speaking, if $\alpha$ is much smaller than $1/e$, 
then the gauge-dependent part becomes of order $e^2$ and can be neglected;
this gives an adequate gauge-parameter region for QED.

There is a difference between QED and the massive vector theory in the
Stueckelberg formalism:
The limit $\alpha\to \infty$ can not be taken in the former case because the third term in
 Eq.~(\ref{eq:HTLQED}) diverges, while
it can in the  latter case.
In this limit, the pole of the fermion propagator in the latter case
 becomes of order $m\ll gT$.

On the other hand, also small $\alpha$ yields the problem; the imaginary part  of the fermion self-energy becomes positive
for $\alpha \ll 1$, which breaks the
 analyticity of the self-energy.

The $\alpha$ dependence of the self-energy can be understood intuitively as follows:
In the Stueckelberg formalism, there are two masses: the physical mass $m$ and 
the unphysical mass $\sqrt{\alpha}m$.
For $gT \gg m$, $\sqrt{\alpha}m$, the self-energy naturally approaches that of QED
because both masses can be neglected.
By contrast, if the unphysical mass $\sqrt{\alpha}m$ is not smaller than $gT$, 
$\sqrt{\alpha}m$ can not be neglected,
although the temperature is enough high compared with the physical mass $m$.
In this case, the self-energy at one loop level does not approach that of QED.

We also note here that in the $g \ll \mOvergT$ case, the boson mass $m$ can not be taken to zero from the outset, but the self-energy is approximately equal to that in the HTL approximation in QED as long as $\alpha \ll 1/g$. 
This implies that the value, $Tg^{3/2}$, is the upper limit of the boson mass that we can neglect.

Summarizing the situation, we see that there are three region of the gauge parameter:
In the first region, the theory approaches QED, and is in adequate gauge parameter region.
In the second region, the theory approaches QED, but is out of adequate gauge parameter region.
In the third region, the theory does  not approach QED, and is out of adequate gauge parameter
 region.
We concludes that the gauge parameter should be chosen to be $1 \lesssim \alpha \ll 1/g$ in
 numerical calculations
in this formalism.

\section{Summary and concluding remarks}
\label{sec:summary}
We have investigated the spectral properties 
 of a fermion coupled with a massive vector boson 
in the  whole temperature ($T$) region at one-loop order.
The vector boson with a mass ($m$) is introduced as a $U(1)$ gauge boson in the
Stueckelberg formalism so that the high $T$ limit, or equivalently the massless limit
in the sense that $m/T\rightarrow 0$, can be taken\footnote{
As is mentioned in Introduction,
the Proca formalism does not yield sensible results for $T \gg m$ 
\cite{proca-problem,kitazawa,dolan-jackiw}.
}:
We have successfully analyzed and clarified the
characteristics of the spectral properties of the fermion in 
the distinct three regions of $T$, i.e.
(I)\,$T \ll m$,\, (II)\,$T \sim m$ and (III)\,$T \gg m$ regions, in a unified way. 
We have also carefully  examined the possible gauge dependence of the spectral properties
of the fermion in the respective three $T$ regions, separately.

In the region (I), the fermion spectral properties hardly change from those in the vacuum,
which are gauge-independent.
In the region (II), the fermion spectral function 
gets to have a three-peak structure in the small momentum region
 with  supports in the positive, zero and negative
energy regions; the three-peak structure becomes prominent when $T\simeq 2m$ for $g=0.5$.
We have confirmed  numerically 
that the fermion poles and hence the fermion spectral function shows virtually no
dependence on the gauge parameter ($\alpha$) for $T \sim m$.
It is thus natural that 
the similar three-peak structure of the fermion spectral function was obtained
 in the Proca formalism for the massive vector field for  $T \sim m$ \cite{kitazawa}, since
the Proca formalism exactly corresponds to the unitary gauge $\alpha \rightarrow \infty$.
Conversely speaking, the three-peak structure 
found in \cite{kitazawa} is not an artifact by a special choice of the gauge and is 
physical.

It is interesting that the spectral function of a fermion coupled with a scalar massive boson
 shows also a similar three-peak structure for $T \sim m$ at the one-loop level 
\cite{kitazawa,mitsutani}.
The present analysis has established that a fermion coupled with a massive boson with a mass
$m$ has  a three-peak structure for small momenta with supports
in the positive, vanishing and negative energy regions at temperatures comparable with
the boson mass, irrespective of the type of the boson, at the one-loop order.

For $T \gg m$\, (region (III)), the fermion spectral function 
tends to have distinct two peaks precisely corresponding to those seen in QED 
in the HTL approximation \cite{frenkel-taylor,braaten-pisarski,weldon:1982aq,weldon}.
It means that our formalism nicely describes the spectral function of the fermion coupled
with a massive vector boson even in the high $T$ region.
There is, however,  a tricky point related to a possible gauge dependence.
We have found that there exists an adequate region of the gauge parameter $\alpha$ in the
high-$T$ region for the perturbation theory at finite $T$:
If $\alpha$ is of the order $1$, the analysis at the one-loop order makes no
problem and is reliable, keeping the positivity of the spectral function and so on;
otherwise, however, these fundamental properties may  be lost.
This is because there exist two mass parameters, i.e., the vector boson mass $m$ and 
the ghost mass $\sqrt{\alpha}m$ inherently in the Stueckelberg formalism.
Thus the precise high-$T$ region should  be defined by the two conditions,
$T \gg m$ {\em and} $T \gg \sqrt{\alpha}m$.
Our extensive analytic study
has proved this observation and 
showed that when $\alpha \ll 1/g$ ($g$ is the coupling constant),
the one-loop analysis is  reliable even in the region (III).
Accordingly, if the unitary gauge ($\alpha \rightarrow \infty$) is adopted for the massive 
vector boson,  the one-loop analysis can not be valid for $T \gg m$, 
as is shown in a different context \cite{proca-problem,dolan-jackiw}.

Our numerical calculation has shown
that there still remains a peak  at the origin in the $\omega$-$|\vp|$ plane, 
though with a faint strength even in the region (III);
this is in contrast to QED  in  the HTL approximation where
such a peak is absent. 
Although it is an interesting possibility that the
three peak structure persists  at the  high-$T$ region and even in QED,
a sensible analysis 
of the spectral properties around such a low-energy region 
requires a resummed perturbation theory to deal with
 possible pinch singularities \cite{pinchSingularity}.
Thus  we leave an analysis of the spectral properties in the very low
energy region as a future work and hope to report elsewhere \cite{persistency}.

We can think of some physical situations where the present analysis can be relevant,
since massive vector bosons at finite $T$ appear in various physical systems.
In QCD, vector bosons or vector-bosonic modes may decrease their masses in association 
with the restoration of chiral symmetry at finite $T$ \cite{rhomeson}.
It would be not surprising if there exist  vector bosonic modes even in the deconfined 
and chiral symmetric phase in the vicinity of the critical temperature $T_c$,
since the existence of other hadronic modes~\cite{kunihiro,detar,shuryak} and bound states~\cite{charmonium} are suggested in that temperature region. 
There may also exist a vector-type glue ball in such a system.
Then the present analysis would suggest that  the quark spectra can be 
largely affected in such a system where the boson mass is comparable to 
$T$ in the order of magnitude.
In the electro-weak theory, the dispersion relation of neutrinos at high $T$ may possibly
be affected by the weak bosons the masses of which change with 
$T$ \cite{boyanovsky}.
One of the findings of the present analysis
tells us that the one-loop analysis in the unitary gauge can not be applicable
when $m/T \ll 1$, which includes  the vicinity of the critical point.

\section*{ACKNOWLEDGMENTS}

We thank M.~Kitazawa and M.~Harada for helpful comments and discussions.
This work was supported by the Grant-in-Aid for the Global COE Program ``The Next Generation of Physics, Spun from Universality and Emergence'' from the Ministry of Education, Culture, Sports, Science and Technology (MEXT) of Japan and by a Grant-in-Aid for Scientific Research by the Ministry of Education, Culture, Sports, Science and Technology (MEXT) of Japan (Nos. 20540265, 19$\cdot$07797).

\appendix

\section{THE CORRESPONDENCE BETWEEN THE STUECKELBERG FORMALISM AND THE ABELIAN HIGGS MODEL}
\label{app:higgs}

In this Appendix, we briefly show that
the abelian Higgs model is reduced to the $U(1)$ gauge theory 
with a massive gauge boson 
in the Stueckelberg formalism~\cite{stueckelberg-review,higgs-stueckelberg}.

The Lagrangian of the abelian Higgs model reads
\begin{align}
{\cal L}_{\text {Higgs}}=-\frac{1}{4}F^{\mu \nu}F_{\mu \nu}+|(\partial _\mu-ieA_\mu)\Phi|^2,
\end{align}
with $F_{\mu\nu}=\partial_{\mu}A_{\nu}-\partial_{\nu}A_{\mu}$.
Here $A_\mu$ and $\Phi$ denote the vector and  the Higgs field, respectively.

We fix the absolute value of the Higgs field $\Phi$ and use the following polar representation:
\begin{align}
\Phi=\frac{m}{e\sqrt2}\exp{\left(\frac{ieB(x)}{m}\right)},
\end{align}
with $|\Phi|=m/(e\sqrt2)$. 
We remark that the scalar field $B$, which will turn to be identified with the
Stueckelberg field,  is introduced as the phase of $\Phi$.
Then, the Lagrangian becomes
\begin{align}
{\cal L}_{\text {Higgs}}=-\frac{1}{4}F^{\mu \nu}F_{\mu \nu}+\frac{m^2}{2}\left(A^\mu-\frac{1}{m}\partial^\mu B\right)^2,
\end{align}
which exactly gives the free Lagrangian
of the massive vector field in the Stueckelberg formalism and
its interaction term with the  Stueckelberg field $B$
given in Eq.~(\ref{eq:formalism-lagrangian}) in the text.
This is what we wanted to show.

\section{CALCULATION OF THE FERMION SELF-ENERGY}
\label{app:self-energy}

Here we derive Eq.~(\ref{eq:stueckelberg-imsigma-exact}) in the text.
We first recall that the retarded self-energy $\Sigma^{\text R}({\mathbf p}, \omega)$
in the one-loop approximation is given from Eq.~(\ref{eq:stueckelberg-selfenergy-tensor})
by the analytic continuation,\, $i\omega_m\rightarrow \omega+i\epsilon \equiv p^0$; 
see Eq.(\ref{eq:retarded-Sigma}).
Then its imaginary part reads
\begin{align}
\begin{split}
{\text {Im}}\Sigma^{\text R}({\mathbf p},\omega)=&-2g^2\gamma^\mu {\text {Im}}\tilde{B}_\mu({\mathbf p},\omega;m)
 +\frac{g^2}{m^2}\Bigl[\not{p}\Bigl(p^2\left({\text {Im}}\tilde{B}({\mathbf p},\omega;\sqrt{\alpha}m)
-{\text {Im}}\tilde{B}({\mathbf p},\omega;m)\right)\\
&-m^2\left(\alpha {\text {Im}}\tilde{B}({\mathbf p},\omega;\sqrt{\alpha}m)-
{\text {Im}}\tilde{B}({\mathbf p},\omega;m)\right)\Bigr) -
p^2\gamma^\mu\left({\text {Im}}\tilde{B}_\mu({\mathbf p},\omega;\sqrt{\alpha} m)-
{\text {Im}}\tilde{B}_\mu({\mathbf p},\omega;m)\right)\Bigr],
\end{split}
\end{align}
where $p^2=\omega^2-|\vp|^2$.
By applying the projection operator for the particle sector,
we have
\begin{align}
\label{eq:app-imsigma}
\begin{split}
{\text {Im}}\Sigma_+({\mathbf p},\omega)=
&\frac{1}{2}\Tr[{\text {Im}}\Sigma^{\text R}({\mathbf p},\omega)\Lambda_{+}({\mathbf p})\gamma^0]\\
=&-2g^2\left({\text {Im}}\tilde{B}^0({\mathbf p},\omega;m)-
\hat{\vp}_i{\text {Im}}\tilde{B}_i({\mathbf p},\omega;m)\right) \\
&+\frac{g^2}{m^2}\Bigl[(\omega-|\vp|)\Bigl(p^2\left({\text {Im}}\tilde{B}({\mathbf p},\omega;\sqrt{\alpha}m)
-{\text {Im}}\tilde{B}({\mathbf p},\omega;m)\right)-
m^2\left(\alpha {\text {Im}}\tilde{B}({\mathbf p},\omega;\sqrt{\alpha}m)-
{\text {Im}}\tilde{B}({\mathbf p},\omega;m)\right)\Bigr) \\
&-p^2\left({\text {Im}}\tilde{B}^0({\mathbf p},\omega;\sqrt{\alpha} m)- 
{\text {Im}}\tilde{B}^0({\mathbf p},\omega; m)- 
\hat{\vp}_i{\text {Im}}\tilde{B}_i({\mathbf p},\omega; \sqrt{\alpha}m)+
\hat{\vp}_i{\text {Im}}\tilde{B}_i({\mathbf p},\omega; m)\right)\Bigr].
\end{split}
\end{align}
Here we have used the relation $\Tr[\Slash{p}\Lambda_{+}({\mathbf p})\gamma^0]/2=\omega-|\vp|$ and 
$\Tr[\tilde{B}^\mu({\mathbf p},\omega;m)\gamma_\mu\Lambda_{+}({\mathbf p})\gamma^0]
=\tilde{B}^0({\mathbf p},\omega;m)-\hat{\vp}_i\tilde{B}_i({\mathbf p},\omega;m)$.
Notice that ${\text {Im}}\tilde{B}({\mathbf p},\omega; m)$ 
and ${\text {Im}}\tilde{B}^\mu({\mathbf p},\omega;m)$ contain 
 $\delta(sE_f-tE_b-\omega)$ with $s,t=\pm 1$, $E_f=|\vect{k}|$, and $E_b=\sqrt{m^2+({\vect{p}-\vect{k}})^2}$,
 while ${\text {Im}}\tilde{B}({\mathbf p},\omega; \sqrt{\alpha}m)$ and 
${\text {Im}}\tilde{B}^\mu({\mathbf p},\omega;\sqrt{\alpha} m)$ 
contain $\delta(sE_f-tE'_b-\omega)$ with $E'_b=\sqrt{\alpha m^2+({\vect{p}-\vect{k}})^2}$.
Collecting the terms which contain $\delta(sE_f-tE_b-\omega)$, we have
\begin{align}
\begin{split}
(\delta(sE_f-tE_b-\omega) {\text {\ term}})=&g^2\int\frac{d^3\vk}{(2\pi)^3}\sum_{s,t=\pm1}\frac{st\pi}{4E_fE_b}\delta(sE_f-tE_b-\omega)(f(sE_f)+n(sE_f-\omega))\\
&\times\left(-2(sE_f-\hat{\vp}\cdot \vk)-\frac{1}{m^2}(\omega-|\vp|)(p^2-m^2)+\frac{1}{m^2}p^2(sE_f-\hat{\vp}\cdot \vk) \right).
\end{split}
\end{align}
Here we have used the following formulae;
\begin{align}
\label{eq:app-imaginary-B}
{\text {Im}}\tilde{B}({\mathbf p},\omega; m)= 
&\int\frac{d^3\vk}{(2\pi)^3}\sum_{s,t=\pm1}\frac{st\pi}{4E_fE_b}\delta(sE_f-tE_b-\omega)(f(sE_f)+n(sE_f-\omega)),\\
\label{eq:app-imaginary-B0}
{\text {Im}}\tilde{B}^0({\mathbf p},\omega; m)= 
&\int\frac{d^3\vk}{(2\pi)^3}\sum_{s,t=\pm1}\frac{st\pi}{4E_fE_b}\delta(sE_f-tE_b-\omega)
(f(sE_f)+n(sE_f-\omega))sE_f,\\
{\text {Im}}\tilde{B}^i({\mathbf p},\omega; m)= 
&\int\frac{d^3\vk}{(2\pi)^3}\sum_{s,t=\pm1}\frac{st\pi}{4E_fE_b}k^i\delta(sE_f-tE_b-\omega)(f(sE_f)+n(sE_f-\omega)).
\end{align}

Using the constraint $\hat{\vp}\cdot \vk=E_f\cos\theta=(m^2-p^2+2\omega sE_f)/2|\vp|$ posed by the delta function, we have
\begin{align}
\label{eq:app-stueckelberg-1}
\begin{split}
(\delta(sE_f-tE_b-\omega) {\text {\ term}})=&g^2\int\frac{d^3\vk}{(2\pi)^3}\sum_{s,t=\pm1}\frac{st\pi}{4E_fE_b}\delta(sE_f-tE_b-\omega)(f(sE_f)+n(sE_f-\omega))\\
&\times\left( \frac{p^2-m^2}{2|\vp|m^2}((\omega-|\vp|)^2-2m^2)+sE_f\frac{2m^2-p^2}{m^2|\vp|}(\omega-|\vp|)\right)\\
=&\frac{g^2}{32\pi m^2|\vp|^2}\int^\infty_0dE_f \sum_{s,t=\pm1}st[(p^2-m^2)((\omega-|\vp|)^2-2m^2)+2sE_f(\omega-|\vp|)(2m^2-p^2)]\\
&\times(f(sE_f)+n(sE_f-\omega))\int^{e_+}_{e_-}dE_b\delta(sE_f-tE_b-\omega),
\end{split}
\end{align}
with $e_{\pm}\equiv\sqrt{(|\vp|\pm E_f)^2+m^2}$. Here the momentum integral is converted to that for the
particle  energies through the formula
\begin{equation}
 \int d^3\vk=\frac{2\pi}{|\vp|}\int^\infty_0 dE_fE_f \int^{e_+}_{e_-}dE_bE_b.
\end{equation}

We now evaluate the integral separately for the time-like and the space-like region.
For the time-like region ($|\omega|>|\vp|$), 
we have 
\begin{align}
\label{eq:app-stueckelberg-timelike}
\begin{split}
(\delta(sE_f-tE_b-\omega) {\text {\ term}})=&\frac{g^2}{32\pi|\mathbf{p}|^2m^2}\int^{E^{-}_{f}}_{E^{+}_{f}}dE_f(f(E_f)+n(E_f-\omega))[(-p^2+m^2)((|\mathbf{p}|-\omega)^2-2m^2) \\
&+2(p^2-2m^2)E_f(\omega-|\mathbf{p}|)],
\end{split}
\end{align}
with $E^\pm_f\equiv (p^2-m^2)/(2(\omega\pm |\vp|))$.
For the space like region ($|\omega|<|\vp|$),
we have
\begin{align}
\label{eq:app-stueckelberg-spacelike}
\begin{split}
(\delta(sE_f-tE_b-\omega) {\text {\ term}})=&\frac{g^2}{32\pi|\mathbf{p}|^2m^2}\int^{E^{-}_{f}}_{E^{+}_{f}}dE_f(f(E_f)+n(E_f-\omega))\\
&\times[(-p^2+m^2)((|\mathbf{p}|-\omega)^2-2m^2) +2(p^2-2m^2)E_f(\omega-|\mathbf{p}|)] \\
&-\frac{g^2}{32\pi|\mathbf{p}|^2m^2}[(p^2-2m^2)(\omega-|\mathbf{p}|)\pi^2 T^2+\omega[2m^4-p^2(|\mathbf{p}|(\omega-|\mathbf{p}|)+m^2)]].
\end{split}
\end{align}
Here we have used the integral formulae
\begin{align}
\int^\infty_{-\infty}dE_f(f(E_f)+n(E_f-\omega)) =&-\omega, \\
\int^\infty_{-\infty}dE_f(f(E_f)+n(E_f-\omega))E_f =& \frac{\pi^2T^2}{2}-\frac{\omega^2}{2}.
\end{align}

Combining Eqs.~(\ref{eq:app-stueckelberg-timelike}) and~(\ref{eq:app-stueckelberg-spacelike}), we have
\begin{align}
\label{eq:app-stueckelberg-unitary}
\begin{split}
(\delta(sE_f-tE_b-\omega) {\text {\ term}})=&\frac{g^2}{32\pi|\mathbf{p}|^2m^2}\int^{E^{-}_{f}}_{E^{+}_{f}}dE_f(f(E_f)+n(E_f-\omega))\\
&\times[(-p^2+m^2)((|\mathbf{p}|-\omega)^2-2m^2) +2(p^2-2m^2)E_f(\omega-|\mathbf{p}|)] \\
&-\frac{g^2}{32\pi|\mathbf{p}|^2m^2}\theta(-p^2)[(p^2-2m^2)(\omega-|\mathbf{p}|)\pi^2 T^2+\omega[2m^4-p^2(|\mathbf{p}|(\omega-|\mathbf{p}|)+m^2)]].
\end{split}
\end{align}

Next, we collect the terms which contain $\delta(sE_f-tE'_b-\omega)$;
\begin{align}
\label{eq:app-stueckelberg-3}
\begin{split}
(\delta(sE_f-tE'_b-\omega) {\text {\ term}})=&\frac{g^2}{32\pi m^2|\vp|^2}\int^\infty_0dE_f \sum_{s,t=\pm1}st
(\omega-|\vp|)[-(\omega-|\vp|)(p^2-\alpha m^2)+2p^2sE_f)]\\
&\times(f(sE_f)+n(sE_f-\omega))\int^{e'_+}_{e'_-}dE'_b\delta(sE_f-tE'_b-\omega)\\
=&-\frac{g^2}{32\pi|\mathbf{p}|^2m^2}\int^{E'^{-}_{f}}_{E'^{+}_{f}}dE_f(f(E_f)+n(E_f-\omega)) [(-p^2+m^2\alpha)(|\mathbf{p}|-\omega)^2+2p^2E_f(\omega-|\mathbf{p}|)]\\
&+\frac{g^2}{32\pi|\mathbf{p}|^2m^2}\theta(-p^2)[p^2(\omega-|\mathbf{p}|)\pi^2 T^2 +\omega(\omega-|\mathbf{p}|)(-\omega p^2-(\omega-|\mathbf{p}|)(-p^2+m^2\alpha))],
\end{split}
\end{align}
where $e'_{\pm}\equiv\sqrt{(|\vp|\pm E_f)^2+\alpha m^2}$ and $E'^\pm_f\equiv (p^2-\alpha m^2)/(2(\omega\pm |\vp|))$.
Combining Eqs.~(\ref{eq:app-stueckelberg-unitary}) and~(\ref{eq:app-stueckelberg-3}), we arrive at
 Eq.~(\ref{eq:stueckelberg-imsigma-exact}).

\section{POWER COUNTING OF THE FERMION SELF-ENERGY}
\label{app:Blimit}

In this appendix, 
we make an order estimate of $\tilde{B}({\mathbf p}, \omega; m)$ and $\tilde{B}^0({\mathbf p}, \omega; m)$ 
used in the analysis done in Sec.~\ref{sec:gauge}.
The estimate will be made separately for the following four limiting cases:
 $T\gg \omega\gg m$, $T\gg m\gg\omega$, $m\gg T \gg\omega$
 and $T\gg \omega\sim m$.
For simplicity, we set $|{\mathbf p}|=0$.

We start with $\tilde{B}(\vz,i\omega_m;m)$ and $\tilde{B}^0(\vz,i\omega_m;m)$ in the imaginary time formalism:
\begin{align}
\label{eq:app-exact-B-st}
\tilde{B}(\vz,i\omega_m;m)=&\int \frac{d^3\vk}{(2\pi)^3}\sum_{s,t=\pm 1}\frac{st}{4E_b E_f}\frac{1}{sE_f-tE_b-i\omega_m}(f(sE_f)+n(tE_b)),\\
\label{eq:app-exact-B0-st}
\tilde{B^0}(\vz,i\omega_m;m)=&\int \frac{d^3\vk}{(2\pi)^3}\sum_{s,t=\pm 1}\frac{st}{4E_b E_f}\frac{sE_ff(sE_f)+(tE_b+i\omega_m)n(tE_b)}{sE_f-tE_b-i\omega_m}.
\end{align}
Performing the analytic continuation and dropping $T=0$ part, 
we get the corresponding retarded functions,
the real parts of which read
\begin{align}
\label{eq:app-real-B-nonst}
\begin{split}
{\text {Re}}\tilde{B}(\vz, \omega;m)\Tnonzero= 
&{\text P}\int \frac{d^3\vk}{(2\pi)^3}\sum_{s,t=\pm 1}\frac{st}{4E_b E_f}
\frac{1}{sE_f-tE_b-\omega}(f(sE_f)+n(tE_b))-\re\tilde{B}(\vz,\omega;m)_{T=0} \\
=& \frac{1}{2\pi^2}{\text P}\int^\infty_0 dE_f \left(E_ff(E_f)\frac{\omega^2-m^2}{(\omega^2-m^2)^2-4E^2_f\omega^2} -n(E_b)\frac{E^2_f}{E_b}\frac{m^2+\omega^2}{(\omega^2-m^2)^2-4E^2_f\omega^2}\right),
\end{split}
\end{align}
\begin{align}
\label{eq:app-real-B0-nonst}
\begin{split}
{\text {Re}}\tilde{B^0}(\vz, \omega; m)\Tnonzero=&{\text P}\int \frac{d^3\vk}{(2\pi)^3}\sum_{s,t=\pm 1}\frac{st}{4E_b E_f}\frac{sE_ff(sE_f)+(tE_b+\omega)n(tE_b)}{sE_f-tE_b-\omega}-\re\tilde{B}^0(\vz,\omega;m)_{T=0} \\
=& \frac{\omega}{2\pi^2}{\text P}\int^\infty_0 dE_f \left(2E^3_ff(E_f)\frac{1}{(\omega^2-m^2)^2-4E^2_f\omega^2} +n(E_b)\frac{E^2_f}{E_b}\frac{2E^2_f+m^2-\omega^2}{(\omega^2-m^2)^2-4E^2_f\omega^2}\right).
\end{split}
\end{align}
Here we have defined $T\neq 0$ parts as $\tilde{B}(\vz,\omega;m)\Tnonzero\equiv\tilde{B}(\vz,\omega;m)-\tilde{B}(\vz,\omega;m)_{T=0}$ and $\tilde{B}^0(\vz,\omega;m)\Tnonzero\equiv\tilde{B}^0(\vz,\omega;m)-\tilde{B}^0(\vz,\omega;m)_{T=0}$.
In the evaluation of these integrals, we have to carefully separate and  
deal with the contributions from the soft region ($E_f \ll T$) and  the hard region ($E_f \sim T$) unlike the calculation of the leading order of the HTL approximation, because the contribution from the soft region can be much larger than that from the hard region in some limits.

From Eqs.~(\ref{eq:app-imaginary-B}) and (\ref{eq:app-imaginary-B0}), the finite temperature parts of $\im \tilde{B}(\vz,\omega;m)$ and $\im \tilde{B}^0(\vz,\omega;m)$ are calculated to be
\begin{align}
\label{eq:app-imaginary-B-nonst}
\begin{split}
{\rm {Im}}\tilde{B}(\vz, \omega; m)\Tnonzero
=&\frac{1}{8\pi}\int^\infty_0 dE_f \frac{E_f}{E_b}
\left[-(f(E_f)+n(E_b))\delta(E_f-E_b+\omega)-(f(E_f)-n(E_b))\delta(E_f+E_b-\omega)\right] \\
=&\frac{\omega^2-m^2}{16\pi\omega^2}\left(\sgn(m^2-\omega^2)f\left(\left|\frac{m^2-\omega^2}{2\omega}\right|\right)+n\left(\frac{m^2+\omega^2}{2\omega}\right)\right),
\end{split}
\end{align}
and
\begin{align}
\label{eq:app-imaginary-B0-nonst}
\begin{split}
{\rm {Im}}\tilde{B}^0(\vz, \omega; m)\Tnonzero
=&-\frac{1}{8\pi}\int^\infty_0 dE_f \frac{E^2_f}{E_b}\left[-(f(E_f)+n(E_b))\delta(E_f-E_b+\omega)+(f(E_f)-n(E_b))\delta(E_f+E_b-\omega)\right] \\
=&\frac{(\omega^2-m^2)^2}{32\pi\omega^3}\left(\sgn(m^2-\omega^2)f\left(\left|\frac{m^2-\omega^2}{2\omega}\right|\right)+n\left(\frac{m^2+\omega^2}{2\omega}\right)\right),
\end{split}
\end{align}
respectively. 
In the first lines, the positiveness of $\omega$ has been taken into account.

\subsection{$T \gg \omega \gg m$}
\label{ssc:app-T-gg-omega-gg-m}
\subsubsection{$\tilde {B}(\vz, \omega; m)$}

We first estimate the imaginary part. 
From Eq.~(\ref{eq:app-imaginary-B-nonst}), we have
\begin{align}
\label{eq:app-T-gg-omega-gg-m-B-im-first}
\begin{split}
{\rm {Im}}\tilde{B}(\vz, \omega; m)\Tnonzero\simeq&
\frac{1}{16\pi}\left(-f\left(\frac{\omega}{2}\right)+
n\left(\frac{\omega}{2}\right)\right)=\frac{1}{8\pi}n(\omega).
\end{split}
\end{align}
Here we have utilized the inequality $\omega \gg m$.
Using the approximate formula
\begin{align}
\label{eq:n-infra}
n(x)\simeq T/x ~{\text {for}}~ x \ll T,
\end{align}
we get
\begin{align}
\label{eq:app-T-gg-omega-gg-m-B-im}
{\rm {Im}}\tilde{B}(\vz, \omega; m)\Tnonzero\simeq&\frac{1}{8\pi}\frac{T}{\omega}.
\end{align}

The real part is given by Eq.~(\ref{eq:app-real-B-nonst}).
To evaluate it, it is found convenient to introduce an intermediate scale $T^*$ 
which satisfies the relation $T \gg T^* \gg \omega$ so that
$T^*$  separates the hard region from the soft region.
The contribution from $E_f>T^*$, which is denoted by $\Bhard$, is evaluated to be
\begin{align}
\begin{split}
\Bhard
\simeq -\frac{1}{8\pi^2}{\text P}\int^\infty_{T^*} dE_f \frac{1}{E_f}\left(f(E_f)-n(E_f)\right).
\end{split}
\end{align}
Here we have used the inequalities
$\omega \gg m$ and $E_f>T^*\gg \omega$ in the hard region.
We get the following expression by integrating by part:
\begin{align}
\label{eq:app-T-gg-omega-gg-m-B-partial-integral}
\begin{split}
\Bhard\simeq&
-\frac{1}{8\pi^2}\left(-\left[f(E_f)-n(E_f)\right]^\infty_{T^*}+{\text P}\int^\infty_{T^*} dE_f \frac{1}{E_f}\frac{d}{dE_f}E_f\left(f(E_f)-n(E_f)\right)\right).
\end{split}
\end{align}
By inserting Eq.~(\ref{eq:n-infra}) and $f(x) \simeq 1/2$ for $x \ll 1$, we see that the first term is of the order of $T/T^*$.
It is easily confirmed that the second term is of the order of $\ln (T^*/T)$ by performing partial integration again, so the second term can be neglected; 
\begin{align}
\label{eq:app-T-gg-omega-gg-m-B-Ef>T*}
\begin{split}
\Bhard\simeq& \frac{1}{8\pi^2}\frac{T}{T^*}.
\end{split}
\end{align}

On the other hand, we have for the contribution from $E_f<T^*$, which is denoted by $\Bsoft$: 
\begin{align}
\begin{split}
\label{eq:app-T-gg-omega-gg-m-B-Ef<T*}
\Bsoft\simeq& -\frac{1}{2\pi^2}{\text P}\int^{T^*}_0 dE_f \frac{TE^2_f}{E^2_f+m^2}\frac{\omega^2}{\omega^4-4E^2_f\omega^2}.
\end{split}
\end{align}
Here again, Eq.~(\ref{eq:n-infra}) has been inserted and the inequality $\omega \gg m$ has been taken into account.
The term which has $f(E_f)$ has been neglected since it is much smaller than $n(E_b)$.
Performing partial fraction decomposition and utilizing the inequality $\omega\gg m$, we get
\begin{align}
\label{eq:app-T-gg-omega-gg-m-B-Ef<T*-2}
\begin{split}
\Bsoft\simeq& -\frac{T}{8\pi^2T^*}+\frac{Tm}{4\pi\omega^2}.
\end{split}
\end{align}
We see that the imaginary part, Eq.~(\ref{eq:app-T-gg-omega-gg-m-B-im}), is much larger than any term of the real part, Eqs.~(\ref{eq:app-T-gg-omega-gg-m-B-Ef>T*}) and~(\ref{eq:app-T-gg-omega-gg-m-B-Ef<T*-2}).

A remark is in order here:\, In the present analysis, we have considered only 
the leading term in the high temperature expansion as in \cite{dolan-jackiw,kapusta},
where the simple approximate formula $n(x)\simeq T/x$ is adopted.
In fact, all the analyses given in this Appendix will be based 
on this formula.
Of course, one could include the next-to-leading term by using the formula $n(x)\simeq T/x -1/2$,
which would lead to logarithmic terms derived in \cite{wang,Mottola}.
The inclusion of such terms is beyond the scope of 
this work.

\subsubsection{$\tilde {B}^0(\vz, \omega;m)$}

The imaginary part of $\tilde {B}^0(\vz, \omega;m)\Tnonzero$ 
is given by Eq.~(\ref{eq:app-imaginary-B0-nonst}), which is estimated to be
\begin{align}
\label{eq:app-T-gg-omega-gg-m-B0-im}
{\rm {Im}}\tilde{B}^0(\vz, \omega; m)\Tnonzero
\simeq\frac{T}{16\pi}.
\end{align}
The derivation of this expression is the same as that of Eq.~(\ref{eq:app-T-gg-omega-gg-m-B-im}).

The real part is given by Eq.~(\ref{eq:app-real-B0-nonst}).
We introduce $T^*$ which satisfies $T \gg T^* \gg \omega$, as before.
The contribution from the $E_f>T^*$ region, which is written as $\Bzerohard$, is evaluated as follows by inserting $E_b\simeq E_f$ and $E_f\gg\omega \gg m$ again:
\begin{align}
\label{eq:app-T-gg-omega-gg-m-B0-Ef-gg-T*}
\begin{split}
\Bzerohard\simeq& \frac{\omega}{4\pi^2}\int^\infty_{T^*} dE_f E_f\frac{-1}{\omega^2}(f(E_f)+n(E_f))\\
=&-\frac{1}{4\pi^2\omega}\left(\int^\infty_{0} dE_f-\int^{T^*}_0dE_f\right) E_f(f(E_f)+n(E_f)).
\end{split}
\end{align}
Inserting Eq.~(\ref{eq:n-infra}), we obtain
\begin{align}
\label{eq:app-T-gg-omega-gg-m-B0-Ef-gg-T*-2}
\begin{split}
\Bzerohard\simeq& -\frac{T^2}{16\omega}+\frac{1}{4\pi^2\omega}\int^{T^*}_0dE_f T\\
=&-\frac{T^2}{16\omega}+\frac{T^* T}{4\pi^2\omega}.
\end{split}
\end{align}

The contribution from $E_f<T^*$ region, represented by $\Bzerosoft$, is estimated as follows by using Eq.~(\ref{eq:n-infra}) and the inequality $\omega \gg m$:
\begin{align}
\begin{split}
\Bzerosoft\simeq& \frac{\omega}{2\pi^2}{\text P}\int^{T^*}_0 dE_f T\frac{E^2_f}{E^2_f+m^2}\frac{2E^2_f-\omega^2}{\omega^4-4E^2_f\omega^2}.
\end{split}
\end{align}
Utilizing the inequalities $T \gg T^* \gg \omega$, $\Bzerosoft$ is evaluated as follows by utilizing partial fraction decomposition:
\begin{align}
\label{eq:app-T-gg-omega-gg-m-B0-Ef<T*}
\begin{split}
\Bzerosoft\simeq& -\frac{TT^*}{4\pi^2\omega}-\frac{T}{4\pi^2\omega}\left(-\pi m +\frac{\omega^2}{4T^*}\right).
\end{split}
\end{align}
Since the largest term is $-T^2/(16\omega)$,  we have
\begin{align}
\label{eq:app-T-gg-omega-gg-m-B0} 
{\rm {Re}}\tilde{B}^0(\vz, \omega;m)\Tnonzero\simeq -\frac{T^2}{16\omega},
\end{align}
which is much larger than the imaginary part.
A few remarks are in order here:~the last formula leads to the well-known result of the HTL approximation in QED.
It is easy to understand if we remember that we can neglect $m$ and apply the HTL approximation in this situation, $T\gg \omega \gg m$.

From Eqs.~(\ref{eq:results-tensor-6}) and~(\ref{eq:results-tensor-7}), it seems that $Tm/(4\pi\omega)$ in Eq.~(\ref{eq:app-T-gg-omega-gg-m-B0-Ef<T*}) yields terms which are of the order of $g^2T\omega/m$ and $g^2T\omega\sqrt{\alpha}/m$ in $\Sigma_+(\vz,\omega)$, which implies that the massless limit ($m/T\rightarrow 0$) can not be taken.
However, from Eq.~(\ref{eq:app-T-gg-omega-gg-m-B-Ef<T*-2}), we see that these terms are canceled by terms which come from Eqs.~(\ref{eq:results-tensor-2}) and~(\ref{eq:results-tensor-3}).

The results of this section, Eqs.~(\ref{eq:app-T-gg-omega-gg-m-B-Ef>T*}), (\ref{eq:app-T-gg-omega-gg-m-B-Ef<T*-2}), (\ref{eq:app-T-gg-omega-gg-m-B0-Ef-gg-T*-2}), and (\ref{eq:app-T-gg-omega-gg-m-B0-Ef<T*}), reproduce leading term of asymptotic form of some integrals used in \cite{Mottola} since the QED limit, $m/T$, $m/\omega\rightarrow 0$, is included in the limit considered in this section, $T\gg\omega \gg m$.
Results in other sections are not related to them because their situations are far from QED.   
\subsection{$T \gg m \gg \omega$}
\subsubsection{$\tilde {B}(\vz, \omega;m)$}

The imaginary part of $\tilde {B}(\vz, \omega;m)\Tnonzero$ 
is given by Eq.~(\ref{eq:app-imaginary-B-nonst}), which is estimated as follows:
\begin{align}
\label{eq:app-T-gg-m-gg-omega-B-im}
\begin{split}
{\rm {Im}}\tilde{B}(\vz, \omega;m)\Tnonzero\simeq& \frac{-m^2}{16\pi\omega^2}\left(f\left(\frac{m^2}{2\omega}\right)+n\left(\frac{m^2}{2\omega}\right)\right)=\frac{-m^2}{8\pi\omega^2}\exp\left({\frac{m^2}{2\omega T}}\right)n\left(\frac{m^2}{\omega}\right)\\
\simeq&\left\{ \begin{array}{ll}
 \frac{-m^2}{8\pi\omega^2}\exp\left(-{\frac{m^2}{2\omega T}}\right) &~(m^2 \gg \omega T~{\textrm {case}}) \\
 -\frac{T}{8\pi\omega} &~(m^2 \ll \omega T~{\textrm {case}}) ~~. \\
\end{array} \right.
\end{split}
\end{align}
In the first line the inequality $m \gg \omega$ has been taken into account,
while in the second line, either $m^2 \gg \omega T$ or $m^2 \ll \omega T$ has been used.

Next, we estimate the real part by  separating the cases  $m^2 \gg \omega T$ and $\omega T \gg m^2$.

\begin{itemize}
\item $m^2 \gg \omega T$ case
\end{itemize}

We begin with Eq.~(\ref{eq:app-real-B-nonst}).
We again introduce an intermediate scale $T^*$ satisfying  $T \gg T^* \gg m$.
The contribution from $E_f>T^*$ is evaluated as follows:
\begin{align}
\label{eq:app-T-gg-m-gg-omega-B-Ef-gg-T*}
\begin{split}
\Bhard\simeq& -\frac{1}{2\pi^2m^2}\int^\infty_{T^*} dE_f E_f(f(E_f)+n(E_f)),
\end{split}
\end{align}
where the relation $E_b\simeq E_f\gg m\gg \omega$ has been utilized and the contribution from $E_f\gtrsim m^2/\omega$ has been neglected because it is suppressed by the Boltzmann factor $\sim \exp(-m^2/(\omega T))$. 
Dividing the interval of the integral as in Eq.~(\ref{eq:app-T-gg-omega-gg-m-B0-Ef-gg-T*}), the following expression is obtained:
\begin{align}
\begin{split}
\Bhard\simeq&-\frac{1}{2\pi^2m^2}\left(\int^\infty_{0} dE_f-\int^{T^*}_0dE_f\right) E_f(f(E_f)+n(E_f))\\
\simeq&-\frac{T^2}{8m^2}+\frac{T^* T}{2\pi^2m^2}.
\end{split}
\end{align}
In the last line, we have inserted Eq.~(\ref{eq:n-infra}).
On the other hand, using Eq.~(\ref{eq:n-infra}) and $m\gg \omega$,
the contribution from $E_f<T^*$ is estimated to be
\begin{align}
\label{eq:app-T-gg-m-gg-omega-B-Ef<T*}
\begin{split}
\Bsoft\simeq& -\frac{m^2 T}{2\pi^2}{\text P}\int^{T^*}_0 dE_f \frac{E^2_f}{E^2_f+m^2}\frac{1}{m^4-4E^2_f\omega^2}.
\end{split}
\end{align}
Utilizing the inequalities $T^*\ll m^2/\omega$ and $m\gg\omega$, we obtain
\begin{align}
\begin{split}
\Bsoft\simeq&-\frac{TT^*}{2\pi^2m^2}+\frac{T}{4\pi m},
\end{split}
\end{align}
which is obtained by partial fraction decomposition.
Taking only the largest term, we have
\begin{align}
\label{eq:app-T-gg-m-gg-omega-B}
{\rm {Re}}\tilde{B}(\vz, \omega;m)\Tnonzero\simeq-\frac{T^2}{8m^2}, 
\end{align}
which dominates the imaginary part.
\begin{itemize}
\item $m^2 \ll \omega T$ case
\end{itemize}

We introduce $T^*$, which is an intermediate scale satisfying $T \gg T^* \gg m^2/\omega$.
From Eq.~(\ref{eq:app-real-B-nonst}), we have for the contribution from $E_f>T^*$
\begin{align}
\begin{split}
\Bhard\simeq& \frac{m^2}{8\pi^2\omega^2}\int^\infty_{T^*} dE_f \frac{1}{E_f}(f(E_f)+n(E_f))\\
\simeq& \frac{m^2T}{8\pi^2\omega^2T^*}.
\end{split}
\end{align}
In the first line, we have used the inequalities $m \gg \omega$ and $E_b\simeq E_f>T^*\gg m^2/\omega$,
while in the second line an approximation similar to that after Eq.~(\ref{eq:app-T-gg-omega-gg-m-B-partial-integral}) has been utilized.

Substituting Eq.~(\ref{eq:n-infra}), the contribution from $E_f<T^*$ is estimated to be
\begin{align}
\begin{split}
\Bsoft\simeq& -\frac{m^2T}{2\pi^2}{\text P}\int^{T^*}_0 dE_f \frac{E^2_f}{E^2_f+m^2}\frac{1}{m^4-4E^2_f\omega^2}\\
\simeq&-\frac{m^2T}{8\pi^2\omega^2T^*} +\frac{T}{4\pi m}.
\end{split}
\end{align}
In the second line the inequality $T^* \gg m^2/\omega$ was used.
Now we see that every term in the real part is much smaller than the imaginary part.
Combining these results, we have
\begin{align} 
\begin{split}
\tilde {B}(\vz, \omega;m)\Tnonzero\simeq\left\{ \begin{array}{ll}
-\frac{1}{8}(\frac{T}{m})^2 & ~(m^2 \gg \omega T~{\textrm {case}}) \\
-\frac{i}{8\pi}\frac{T}{\omega} & ~(m^2 \ll \omega T~{\textrm {case}}) ~~. \\
\end{array} \right.
\end{split} \end{align} 
\subsubsection{$\tilde {B}^0(\vz, \omega;m)$}

First, let us estimate the imaginary part of $\tilde {B}^0(\vz, \omega;m)\Tnonzero$ 
 given by Eq.~(\ref{eq:app-imaginary-B0-nonst}):
\begin{align}
\label{eq:app-T-gg-m-gg-omega-B0-im}
\begin{split}
{\rm {Im}}\tilde{B}^0(\vz, \omega;m)\Tnonzero
\simeq&\left\{ \begin{array}{ll}
 \frac{m^4}{16\pi\omega^3}\exp\left(-{\frac{m^2}{2\omega T}}\right) &~(m^2 \gg \omega T~{\textrm {case}}) \\
 \frac{m^2T}{16\pi\omega^2} &~(m^2 \ll \omega T~{\textrm {case}}) ~~. \\
\end{array} \right.
\end{split}
\end{align}
We have arrived at this expression in the same way as in Eq.~(\ref{eq:app-T-gg-m-gg-omega-B-im}).

\begin{itemize} 
\item $m^2 \gg \omega T$ case
\end{itemize}

In this case, the real part of $\tilde {B}^0({\mathbf p}, \omega; m)\Tnonzero$ which is given by Eq.~(\ref{eq:app-real-B0-nonst}) is evaluated as follows.
Using an intermediate scale $T^*$ satisfying $T \gg T^* \gg m$,
 we have for the contribution from $E_f>T^*$
\begin{align}
\label{eq:B}
\begin{split}
\Bzerohard\simeq& \frac{\omega}{\pi^2m^4}\int^\infty_{T^*} dE_f E^3_f(f(E_f)+n(E_f)),
\end{split}
\end{align}
where the contribution from $E_f \gtrsim m^2/\omega$ has been neglected since it is suppressed by the Boltzmann factor $\sim \exp(-m^2/(\omega T))$.
Also the inequality $m\gg \omega$ has been taken into account.
Dividing the interval of the integral as in Eq.~(\ref{eq:app-T-gg-omega-gg-m-B0-Ef-gg-T*}), we obtain
\begin{align}
\label{eq:C}
\begin{split}
\Bzerohard\simeq&\frac{\omega}{\pi^2m^4}\left(\int^\infty_{0} dE_f-\int^{T^*}_0dE_f\right) E^3_f(f(E_f)+n(E_f))\\
\simeq&\frac{\pi^2}{8}\frac{\omega T^4}{m^4}-\frac{\omega TT^{*3}}{3\pi^2m^4}.
\end{split}
\end{align}
In the last line, we have substituted Eq.~(\ref{eq:n-infra}).

Next, we estimate the contribution from $E_f<T^*$.
Since the leading terms cancel out, we retain $\omega^2$ in Eq.~(\ref{eq:app-real-B0-nonst}).
If we define $\Omega_{\pm}^2 \equiv \omega^2\pm m^2$, the contribution from $E_f<T^*$ is estimated as follows:
\begin{align}
\begin{split}
\Bzerosoft\simeq& \frac{\omega T}{2\pi^2}{\text P}\int^{T^*}_0 dE_f \frac{E^2_f}{E^2_f+m^2}\frac{2E^2_f-\Omega_{-}^2}{\Omega_{-}^4-4E^2_f\omega^2},
\end{split}
\end{align}
where we have used Eq.~(\ref{eq:n-infra}).
Performing partial fraction decomposition, we have
\begin{align}
\begin{split}
\Bzerosoft\simeq&-\frac{T}{4\pi^2\omega}{\text P}\int^{T^*}_0 dE_f \left(1-\frac{m^2}{E^2_f+m^2}\frac{2\omega^2}{\Omega_{+}^2}-\frac{\Omega_{-}^6}{4\omega^2\Omega_{+}^2}\frac{1}{E^2_f-\Omega_{-}^4/(4\omega^2)}\right)\\
\simeq&-\frac{T}{4\pi^2\omega}\left(-\frac{\pi\omega^2}{m}+T^*\frac{2\omega^2}{\Omega^2_+}+\frac{4}{3}\frac{\omega^2T^{*3}}{\Omega^2_+\Omega^2_-}\right).
\end{split}
\end{align}
In the second line the inequality $T^* \gg m$ has been utilized.
Notice that because of canceling out of the leading terms, we have to retain the next-to-leading term in the expansion of $\ln(1+2\omega T^*/\Omega^2_-)$.
The largest term is the first term of Eq.~(\ref{eq:C}), so
\begin{align}
{\text {Re}}\tilde{B}^0(\vz, \omega;m)\Tnonzero\simeq \frac{\pi^2}{8}\frac{\omega T^4}{m^4},
\end{align}
which is found to be much larger than the imaginary part.
\begin{itemize}
\item $m^2 \ll \omega T$ case
\end{itemize}

The real part is given by Eq.~(\ref{eq:app-real-B0-nonst}).
As is before, we introduce an intermediate scale $T^*$ satisfying  $T \gg T^* \gg m^2/\omega \gg m$.
Then
\begin{align}
\begin{split}
\Bzerohard\simeq&-\frac{T^2}{16\omega}+\frac{T^* T}{4\pi^2\omega}.
\end{split}
\end{align}
This expression has been obtained in the same procedure as in Eqs.~(\ref{eq:app-T-gg-omega-gg-m-B0-Ef-gg-T*}) and~(\ref{eq:app-T-gg-omega-gg-m-B0-Ef-gg-T*-2}).

The contribution from $E_f<T^*$ is estimated as follows.
Using Eq.~(\ref{eq:n-infra}) and the inequality $m \gg \omega$, we have
\begin{align}
\begin{split}
\Bzerosoft\simeq& \frac{\omega}{2\pi^2}{\text P}\int^{T^*}_0 dE_f T\frac{E^2_f}{E^2_f+m^2}\frac{2E^2_f+m^2}{m^4-4E^2_f\omega^2}\\
\simeq& -\frac{TT^*}{4\pi^2\omega}+\frac{m^4T}{16\pi^2\omega^3T^*}.
\end{split}
\end{align}
In the last line we have utilized the inequality $T^* \gg m^2/\omega \gg m$.
Since $T^* \gg m^4/(\omega^2 T)$, the largest term is found to be $-T^2/(16\omega)$, so we have 
\begin{align}
\label{eq:app-T-gg-m-gg-omega-B0-m^2-ll-omegaT} 
{\rm {Re}}\tilde{B}^0(\vz, \omega;m)\Tnonzero\simeq -\frac{T^2}{16\omega}.
\end{align}
This is much larger than the imaginary part.

Combining these results, we have
\begin{align} 
\begin{split}
\tilde {B}^0(\vz, \omega;m)\Tnonzero \simeq \left\{ \begin{array}{ll}
\frac{\pi^2}{8}\frac{\omega T^4}{m^4} & ~(m^2 \gg \omega T~{\textrm {case}}) \\
-\frac{T^2}{16\omega} & ~(m^2 \ll \omega T~{\textrm {case}}) ~~.\\
\end{array} \right.
\end{split} \end{align} 
\subsection{$m \gg T \gg \omega$}
\label{ssc:m-gg-T-gg-omega}
\subsubsection{$\tilde {B}(\vz, \omega;m)$}

We start with an estimate of the  imaginary part of $\tilde {B}(\vz, \omega;m)\Tnonzero$ 
 given by Eq.~(\ref{eq:app-imaginary-B-nonst}).
For $m \gg T \gg \omega$, we have
\begin{align}
\label{eq:app-m-gg-T-gg-omega-B-im}
{\rm {Im}}\tilde{B}(\vz, \omega;m)\Tnonzero\simeq -\frac{m^2}{8\pi\omega^2}\exp\left(-\frac{m^2}{2\omega T}\right).
\end{align}
The derivation of this expression is almost the same as that of Eq.~(\ref{eq:app-T-gg-m-gg-omega-B-im}) in the $m^2\gg\omega T$ case.
The real part of $\tilde {B}(\vz, \omega;m)\Tnonzero$ given by Eq.~(\ref{eq:app-real-B-nonst})
is estimated as follows:
\begin{align}
{\rm {Re}}\tilde{B}(\vz, \omega;m)\Tnonzero
\simeq \frac{1}{2\pi^2}{\text P}\int^\infty_0 dE_f E_ff(E_f)\frac{-m^2}{m^4-4E^2_f\omega^2}.
\end{align}
Here we have used the inequality $m \gg \omega$ and
 $n(E_b)$ has been neglected because of the suppression factor by $\exp(-m/T)$.
Notice that the contribution from $E_f \gtrsim m^2/\omega \gg T$ is negligible 
owing to the suppression factor by $f(E_f)$.
Then, we see that the denominator of the integrand can be approximated as $m^4-4E^2_f\omega^2\simeq m^4$.
Thus we have
\begin{align}
\label{eq:app-m-gg-T-gg-omega-B}
{\rm {Re}}\tilde{B}(\vz, \omega;m)\Tnonzero
\simeq \frac{-1}{2\pi^2m^2}{\text P}\int^\infty_0 dE_f E_ff(E_f) =-\frac{T^2}{24m^2},
\end{align}
which is found to dominate the imaginary part.
\subsubsection{$\tilde {B}^0(\vz, \omega;m)$}

We estimate the imaginary part of $\tilde {B}^0(\vz, \omega;m)\Tnonzero$, which is 
given by Eq.~(\ref{eq:app-imaginary-B0-nonst}).
Utilizing the same procedure as in Eq.~(\ref{eq:app-m-gg-T-gg-omega-B-im}), we have
\begin{align}
{\rm {Im}}\tilde{B}^0(\vz, \omega;m)\Tnonzero
\simeq \frac{m^4}{16\pi\omega^3}\exp\left(-\frac{m^2}{2\omega T}\right).
\end{align}

The real part, which is given by Eq.~(\ref{eq:app-real-B0-nonst}), is estimated as follows:
\begin{align} \begin{split}
{\text {Re}}\tilde{B}^0(\vz, \omega;m)\Tnonzero
\simeq \frac{\omega}{\pi^2}{\text P}\int^\infty_0 dE_f E^3_ff(E_f)\frac{1}{m^4-4E^2_f\omega^2}.
\end{split} \end{align}
Here we have utilized the inequality $m \gg \omega$ and neglected $n(E_b)$ since its contribution is suppressed by the factor $\exp(-m/T)$.
Following the procedure used in obtaining Eq.~(\ref{eq:app-m-gg-T-gg-omega-B}), we have
\begin{align} \begin{split}
{\text {Re}}\tilde{B}^0(\vz, \omega;m)\Tnonzero\simeq& \frac{\omega}{\pi^2 m^4}\int^\infty_0 dE_f E^3_ff(E_f)
=\frac{7\pi^2}{120}\frac{\omega T^4}{m^4}.
\end{split} 
\end{align}
One sees that this term dominates ${\text {Im}}\tilde{B}^0(\vz, \omega;m)\Tnonzero$.

\subsection{$T \gg \omega \sim m$}
\subsubsection{$\tilde {B}(\vz, \omega;m)$}

We start with an estimate of the imaginary part of $\tilde {B}(\vz, \omega;m)\Tnonzero$,
which is given by Eq.~(\ref{eq:app-imaginary-B-nonst}).
Utilizing the inequality $T \gg \omega \sim m$ and substituting Eq.~(\ref{eq:n-infra}), we have
\begin{align} 
\label{eq:app-T-gg-omega-sim-m-B-im}
\begin{split}
{\rm {Im}}\tilde{B}(\vz, \omega;m)\Tnonzero
\simeq\frac{T}{8\pi\omega}\frac{\omega^2-m^2}{\omega^2+m^2}
\sim \frac{T}{\omega}.
\end{split} \end{align}

In the evaluation of the real part, we introduce $T^*$ satisfying $T \gg T^* \gg \omega$.
From Eq.~(\ref{eq:app-real-B-nonst}), we arrive at the following expression by using the inequalities $E_b\sim E_f\gg\omega$, $m$:
\begin{align}
\begin{split}
\Bhard
\simeq -\frac{1}{8\pi^2\omega^2}
{\text P}\int^\infty_{T^*} dE_f \frac{1}{E_f}\left(\Omega_{-}^2f(E_f)-\Omega_{+}^2n(E_f)\right)\simeq 
\frac{\Omega_{+}^2}{8\pi^2\omega^2}\frac{T}{T^*},
\end{split}
\end{align}
with $\Omega_{\pm}^2 \equiv \omega^2\pm m^2$. 
In the last line we have performed the partial integration and the order estimate as we did for Eq.~(\ref{eq:app-T-gg-omega-gg-m-B-partial-integral}).

Inserting Eq.~(\ref{eq:n-infra}), the contribution from $E_f<T^*$ is estimated to be
\begin{align}
\begin{split}
\Bsoft\simeq& \frac{T\Omega^2_+}{8\pi^2\omega^2}{\text P}\int^{T^*}_0 dE_f \frac{E^2_f}{E^2_f+m^2}
\frac{1}{E^2_f-\Omega_{-}^4/(4\omega^2)},
\end{split}
\end{align}
Using the inequality $T\gg T^*\gg \omega\sim m$ and carrying out partial fraction decomposition, the following estimation is obtained:
\begin{align}
\begin{split}
\Bsoft\simeq&-\frac{T}{8\pi^2\omega^2}\frac{\Omega_{-}^4}{\Omega_{+}^2T^*}+\frac{m}{\Omega_{+}^2}\frac{T}{4\pi}
\sim\frac{T}{\omega}.
\end{split}
\end{align}
We see that both of the real part and the imaginary part are of the order of $T/\omega$;
\begin{align}
\label{eq:app-T-gg-omega-sim-m-B}
\tilde{B}(\vz, \omega;m)\Tnonzero\sim\frac{T}{\omega}.
\end{align}

\subsubsection{$\tilde {B}^0(\vz, \omega;m)$}

We make an order estimate of the imaginary part of $\tilde {B}^0(\vz, \omega;m)\Tnonzero$ 
 given by Eq.~(\ref{eq:app-imaginary-B0-nonst}).
Repeating the same derivation leading to Eq.~(\ref{eq:app-T-gg-omega-sim-m-B-im}), we have
\begin{align} \begin{split}
{\rm {Im}}\tilde{B}^0(\vz, \omega;m)\Tnonzero
\simeq\frac{T}{16\pi\omega^2}\frac{(m^2-\omega^2)^2}{m^2+\omega^2}
\sim T.
\end{split} \end{align}

From Eq.~(\ref{eq:app-real-B0-nonst}), we estimate the 
contribution from the $E_f>T^*$ region to its real part as follows:
\begin{align}
\label{eq:app-T-gg-omega-sim-m-B0-Ef>T^*}
\begin{split}
\Bzerohard\simeq -\frac{T^2}{16\omega}+\frac{T^* T}{4\pi^2\omega}.
\end{split}
\end{align}
Here we have introduced $T^*$ which satisfies $T \gg T^* \gg m$,
 and used the same approximations as that adopted in Eqs.~(\ref{eq:app-T-gg-omega-gg-m-B0-Ef-gg-T*}) and~(\ref{eq:app-T-gg-omega-gg-m-B0-Ef-gg-T*-2}).

On the other hand, the contribution from $E_f<T^*$ is estimated as follows by using Eq.~(\ref{eq:n-infra}):
\begin{align}
\begin{split}
\Bzerosoft\simeq& \frac{\omega}{2\pi^2}{\text P}\int^{T^*}_0 dE_f T\frac{E^2_f}{E^2_f+m^2}\frac{2E^2_f-\Omega_{-}^2}{\Omega_{-}^4-4E^2_f\omega^2}\\
\simeq&-\frac{T}{4\pi^2\omega}\Biggl(T^*-m\frac{\pi}{2}+\frac{\Omega_{-}^2}{2}\left(\frac{\Omega_{-}^2}{2\omega^2}-1\right)\left(-\frac{1}{T^*}\right)+m^2\Omega_{-}^2\frac{\Omega_{-}^2-2\omega^2}{\Omega_{-}^4+4m^2\omega^2}\left(m\frac{\pi}{2}+\frac{1}{T^*}\right)\Biggr),
\end{split}
\end{align}
with $\Omega_{-}^2 \equiv \omega^2-m^2$.
In the last line, the inequality $T^* \gg \omega\sim m$ has been utilized and partial fraction decomposition has been performed.
Since the largest term is $-T^2/(16\omega)$ in Eq.~(\ref{eq:app-T-gg-omega-sim-m-B0-Ef>T^*}), we have
\begin{align}
\label{eq:app-T-gg-omega-sim-m-B0}
{\text {Re}}\tilde{B}^0(\vz, \omega;m)\Tnonzero\simeq -\frac{T^2}{16\omega},
\end{align}
which is much larger than the imaginary part.


\begin{thebibliography}{99}
 \bibitem{frenkel-taylor}
 J.~Frenkel and J.~C.~Taylor,
  Nucl.\ Phys.\  B {\bf 334}, 199 (1990).
\bibitem{braaten-pisarski}
 E.~Braaten and R.~D.~Pisarski,
 Nucl.\ Phys.\  B {\bf 337}, 569 (1990);
 {\bf 339}, 310 (1990).

\bibitem{weldon:1982aq}
   H.~A.~Weldon,
  Phys.\ Rev.\  D {\bf 26}, 1394 (1982).

\bibitem{weldon}
 H.~A.~Weldon,
  Phys.\ Rev.\  D {\bf 26}, 2789 (1982); 
{\bf 40}, 2410 (1989).
\bibitem{kitazawa-NJL}
 M.~Kitazawa, T.~Kunihiro and Y.~Nemoto,
  Phys.\ Lett.\  B {\bf 633}, 269 (2006).
  \bibitem{NJL}
 Y.~Nambu and G.~Jona-Lasinio,
 Phys.\ Rev.\  {\bf 122}, 345 (1961);
 Phys.\ Rev.\  {\bf 124}, 246 (1961);
 As a review article, see T.~Hatsuda and T.~Kunihiro,
  Phys.\ Rept.\  {\bf 247}, 221 (1994).
  \bibitem{kunihiro}
 T.~Hatsuda and T.~Kunihiro,  Phys.\ Rev.\ Lett.\  {\bf 55}, 158 (1985).
  \bibitem{kitazawa}
 M.~Kitazawa, T.~Kunihiro and Y.~Nemoto,  Prog.\ Theor.\ Phys.\  {\bf 117}, 103 (2007).
\bibitem{mitsutani}
 M.~Kitazawa, T.~Kunihiro, K.~Mitsutani and Y.~Nemoto, Phys.\ Rev.\  D {\bf 77}, 045034 (2008).
%
\bibitem{stueckelberg}
 E. C. G. Stueckelberg, Helv. Phys. Acta 11, 225 (1938);
 Helv. Phys. Acta 11, 299 (1938);
 312 (1938).
 \bibitem{stueckelberg-review}
 As a review article, see H.~Ruegg and M.~Ruiz-Altaba,  Int.\ J.\ Mod.\ Phys.\  A {\bf 19}, 3265 (2004).

\bibitem{rebhan}
  R.~Kobes, G.~Kunstatter and A.~Rebhan,
  Phys.\ Rev.\ Lett.\  {\bf 64}, 2992 (1990);
  Nucl.\ Phys.\  B {\bf 355}, 1 (1991);
  A.~Rebhan,
  Lect.\ Notes Phys.\  {\bf 583}, 161 (2002).
\bibitem{lebellac}
 M. Le Bellac, ``Thermal Field Theory,'' {\it Cambridge, UK: Univ. Pr. }(1996).
\bibitem{kapusta}
 J.~I.~Kapusta and C.~Gale,　``Finite-temperature field theory: Principles and applications,''
{\it  Cambridge, UK: Univ. Pr.  }(2006).
\bibitem{higgs-stueckelberg}
 T. W. Kibble, in {\it Proc. Int. Conf. on High Energy Physics} (Oxford University Press, 1965).
\bibitem{proca-problem}
  S.~Weinberg,
  Phys.\ Rev.\  D {\bf 9}, 3357 (1974).
\bibitem{dolan-jackiw}
 L.~Dolan and R.~Jackiw,
  Phys.\ Rev.\  D {\bf 9}, 3320 (1974).

\bibitem{pinchSingularity}
 V.~V.~Lebedev and A.~V.~Smilga,
  Annals Phys.\  {\bf 202}, 229 (1990); see also
   S.~Jeon,
  Phys.\ Rev.\  D {\bf 52}, 3591 (1995);
  J.~S.~Gagnon and S.~Jeon,
  Phys.\ Rev.\  D {\bf 75}, 025014 (2007)
  [Erratum-{\it ibid}.\  {\bf 76}, 089902 (2007)];
 J.~S.~Gagnon and S.~Jeon,
  {\it ibid}.\ {\bf 76}, 105019 (2007);
Y.~Hidaka and T.~Kunihiro,  arXiv:1009.5154 [hep-ph].
\bibitem{persistency}
Y.~Hidaka, D.~Satow, and T.~Kunihiro, in progress.
\bibitem{wang}
 S.~Y.~Wang,
  Phys.\ Rev.\  D {\bf 70}, 065011 (2004).
\bibitem{Mottola}
  E.~Mottola and Z.~Szep,
  Phys.\ Rev.\  D {\bf 81}, 025014 (2010).

\bibitem{rhomeson}
 G.~E.~Brown and M.~Rho, Phys.\ Rev.\ Lett.\  {\bf 66}, 2720 (1991);\,
   T.~Kunihiro,
  Nucl.\ Phys.\  B {\bf 351}, 593 (1991);\,
T.~Hatsuda, Y.~Koike and S.~H.~Lee,
  Nucl.\ Phys.\  B {\bf 394}, 221 (1993);\,
 M.~Harada and K.~Yamawaki,  Phys.\ Rev.\ Lett.\  {\bf 86}, 757 (2001);\,
 Y.~Hidaka, O.~Morimatsu and M.~Ohtani,
 Phys.\ Rev.\  D {\bf 73}, 036004 (2006);
 M.~Harada and C.~Sasaki,
 Phys.\ Lett.\  B {\bf 537}, 280 (2002);
 Phys.\ Rev.\  D {\bf 73}, 036001 (2006).

\bibitem{detar}
 C.~E.~Detar,
  Phys.\ Rev.\  D {\bf 32}, 276 (1985);
  C.~E.~Detar and J.~B.~Kogut,
  Phys.\ Rev.\ Lett.\  {\bf 59}, 399 (1987).
\bibitem{shuryak}
  E.~V.~Shuryak and I.~Zahed,
  Phys.\ Rev.\  C {\bf 70}, 021901 (2004).
\bibitem{charmonium}
 S.~Datta, F.~Karsch, P.~Petreczky and I.~Wetzorke,  
 Nucl.\ Phys.\ Proc.\ Suppl.\  {\bf 119}, 487 (2003);
 Phys.\ Rev.\  D {\bf 69}, 094507 (2004).
 T.~Umeda, H.~Matsufuru, O.~Miyamura and K.~Nomura,
  Nucl.\ Phys.\  A {\bf 721}, 922 (2003);
 T.~Umeda, K.~Nomura and H.~Matsufuru,
  Eur.\ Phys.\ J.\  C {\bf 39S1}, 9 (2005);
 M.~Asakawa and T.~Hatsuda,  Phys.\ Rev.\ Lett.\  {\bf 92}, 012001 (2004);

\bibitem{boyanovsky}
D.~Notzold and G.~Raffelt,
  Nucl.\ Phys.\  B {\bf 307}, 924 (1988);
G.~F.~ Giudice, A.~ Notari, M.~ Raidal, A.~ Riotto, A.~ Strumia,
  {\it ibid}.\ {\bf 685}, 89 (2004);\,
 D.~Boyanovsky, Phys.\ Rev.\  D {\bf 72}, 033004 (2005). 

\end{thebibliography}
\end{document}